\batchmode
\makeatletter
\def\input@path{{/CTA//}}
\makeatother
\documentclass[oneside,english]{book}
\usepackage[T1]{fontenc}
\usepackage[latin9]{inputenc}
\usepackage[a4paper]{geometry}
\usepackage{babel}

\usepackage{array}
\usepackage{varioref}
\usepackage{float}
\usepackage{rotfloat}
\usepackage{amsthm}
\usepackage{amsmath}
\usepackage{graphicx}
\usepackage[authoryear]{natbib}
\usepackage[unicode=true, 
 bookmarks=false,
 breaklinks=false,pdfborder={0 0 0},backref=false,colorlinks=false]
 {hyperref}
\hypersetup{pdftitle={Southern Africa CTA sites},
 pdfauthor={Dr. PP Kriger}}
 
\makeatletter

\providecommand{\tabularnewline}{\\}

\usepackage{endnotes}
\let\footnote=\endnote

\@ifundefined{showcaptionsetup}{}{%
 \PassOptionsToPackage{caption=false}{subfig}}
\usepackage{subfig}
\makeatother

\begin{document}
\noindent \begin{flushright}
\textbf{\thispagestyle{empty}\includegraphics[bb=200bp 0bp 468bp 96bp,clip,width=0.3\columnwidth]{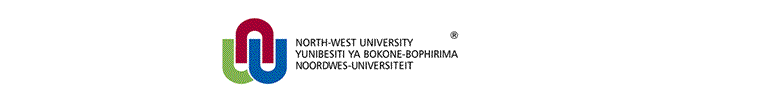}}
\par\end{flushright}

\begin{center}
\textbf{\Huge ~}\\
\textbf{\Huge ~}\\
\textbf{\Huge Southern Africa CTA Site Proposal}\\
\textbf{\Huge ~}
\par\end{center}{\Huge \par}

\begin{center}
P.P. Krüger and D.J. van der Walt\\
North-West University, South Africa
\par\end{center}

\begin{center}
June 2011
\par\end{center}

\textbf{\huge Summary}{\huge \par}

~\\
Southern Africa has some of the world's best sites for air Cherenkov
telescopes. South Africa has only one viable site, which is south
of Sutherland and also close to the Southern African Large Telescope
(SALT). This site has very good infrastructure and is easy to access,
but only 47\% of the night-time has a cloudless sky usable for observations.

Namibia, which already hosts the H.E.S.S telescope, has a number of
potential sites with much less cloud coverage. The H.E.S.S.~site
is one of the highest of these sites at 1840~m~a.s.l. with about
64\% of the night-time cloudless. It also has very low night sky background
levels and is relatively close (about 100~km) to Windhoek. Moving
further away from Windhoek to the south, the cloud coverage and artificial
night sky brightness becomes even less, with the site at Kuibis (between
Keetmanshoop and Luderitz) at 1640~m~a.s.l. having clear night skies
73\% of the time. Even though this site seems remote (being 660~km
from Windhoek by road), it is close to the national B4 highway, a
railway line, a power line and an optical fiber line. It is also less
than two hours drive away from a harbour and national airports. The
Namibian sites also receive very little snow, if any, and the wind
speeds are less than 50~km/h for more than 90\% of the time with
maximum wind speeds of around 100~km/h. Seismically the whole Southern
African region is very stable.

In view of the geographic advantage of Southern Africa for astronomy,
the South African government is committed to multi-wavelength astronomy.
The region around SALT and MeerKAT in the Northern Cape has been declared
a protected region for astronomy and the government has already invested
relatively large amounts of money into the SALT and MeerKAT/SKA projects.
The government is also committed to uplifting the whole of Southern
Africa and would support a proposed CTA site in Namibia. A number
of research chair grants have been allocated to various universities,
with the North-West University holding a chair in high-energy astrophysics
to stimulate growth in gamma-ray astronomy. Currently, it is under
the auspices of this chair that South Africa is part of the H.E.S.S.~and
CTA collaborations.

\begin{figure}[h]
\noindent \begin{centering}
\includegraphics[width=0.47\columnwidth]{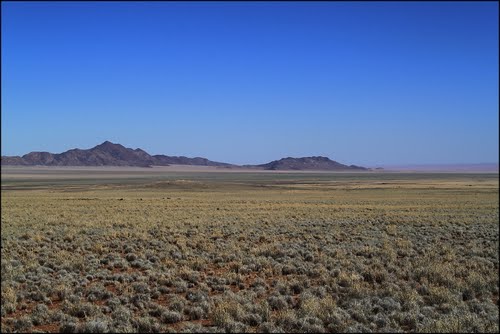}~\includegraphics[width=0.47\columnwidth]{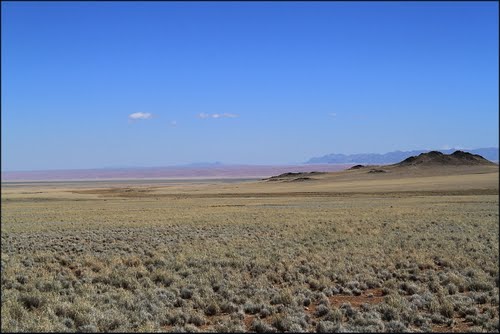}
\par\end{centering}

\caption{Photo taken at Kuibis (region 2) in Namibia.\label{fig:Photo-Site2}}

\end{figure}

\chapter{Overview of Southern Africa}

First the main CTA site specifications \citep{CTA2010} of altitude,
cloud coverage, flatness, night sky background and seismic events
will be considered to evaluate potential CTA sites in Southern Africa.
Thereafter, the economic and political conditions of South Africa
and Namibia will be considered. The best sites in South Africa and
Namibia, apart from the H.E.S.S site, will then be further investigated
in Chapters 2 and 3.

\section{Height and cloud coverage}

\begin{table}
\caption{Table of average night-time cloud coverage and percentage of clear
night skies for the regions shown in Figure \ref{fig:Contour-map},
calculated using the MODIS satellite data. \label{tab:Table-satellite}}

\begin{tabular}{|c|c|c|c|c|c|}
\hline 
Site &  & \multicolumn{1}{c|}{1) Sutherland} & 2) Kuibis & 3) Maltahohe & 4) H.E.S.S.\tabularnewline
\hline 
Height  & m a.s.l & \multicolumn{1}{c|}{1560} & 1640 & 1700 & 1840\tabularnewline
\hline 
Cloud cover & \% & 32.4$\pm1.1$ & 13.3$\pm0.7$ & 18.0$\pm0.9$ & 21.2$\pm1.1$\tabularnewline
\hline 
Time clear sky & \% & 46.5$\pm1.0$ & 73.0$\pm0.7$ & 70.3$\pm1.1$ & 64.4$\pm1.4$\tabularnewline
\hline
\end{tabular}
\end{table}

\begin{figure}[b]
\begin{centering}
\includegraphics[bb=0bp 0bp 900bp 1050bp,scale=0.25]{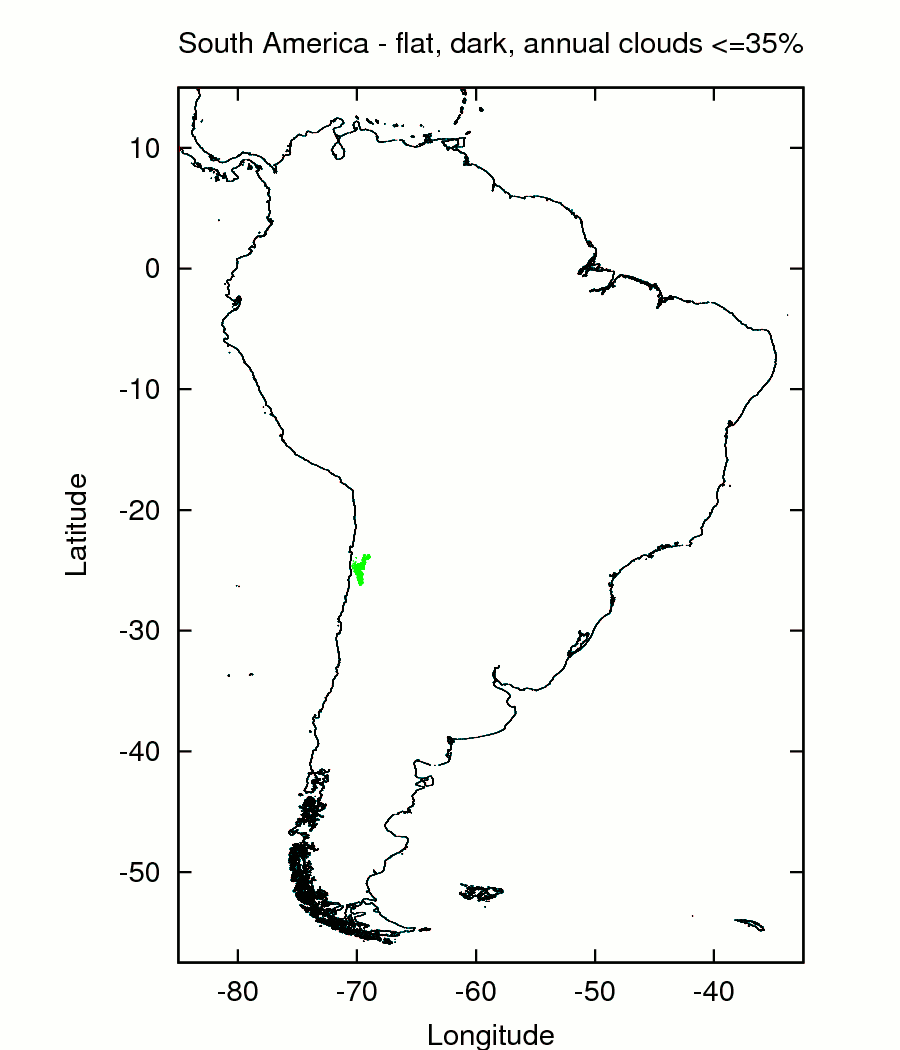}\includegraphics[scale=0.25]{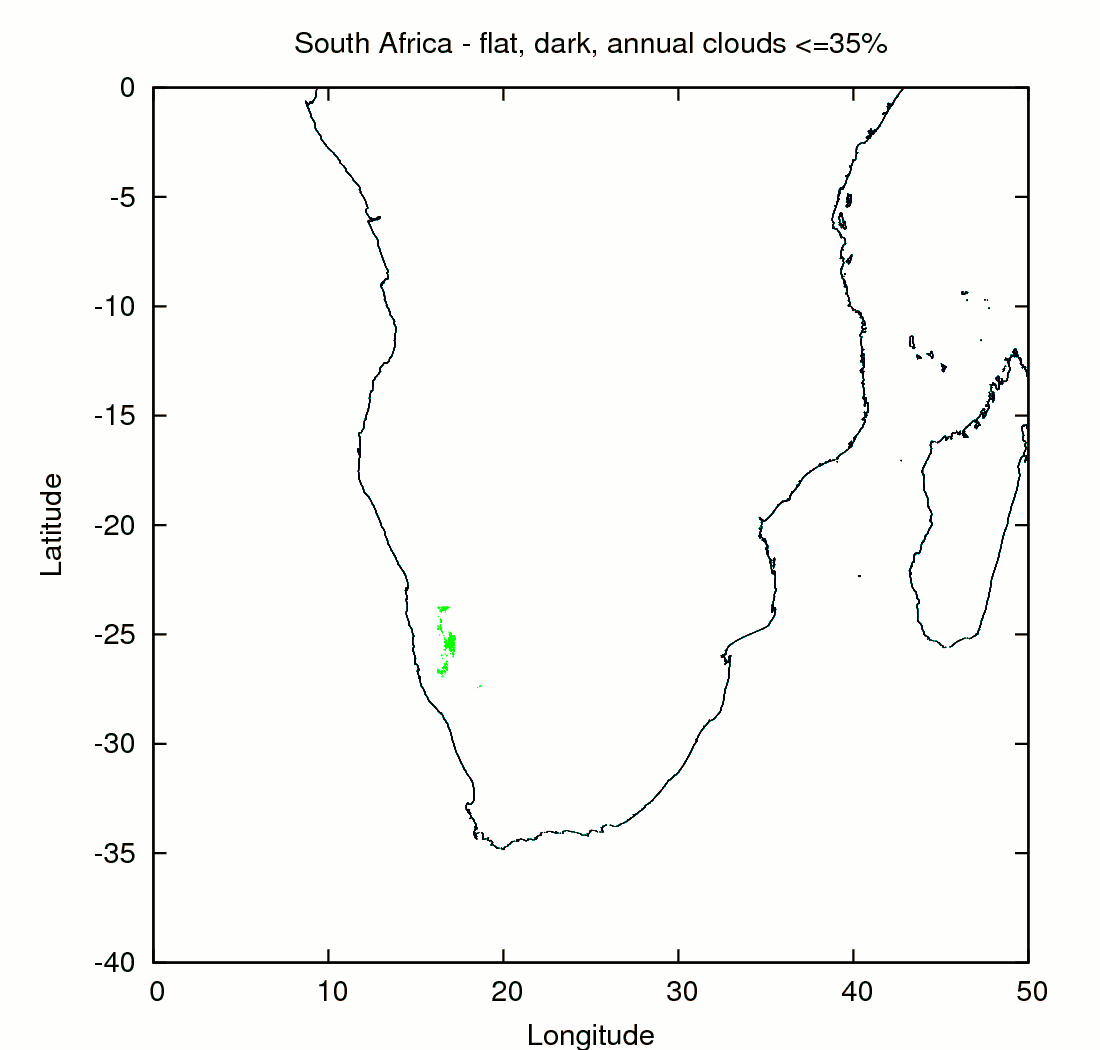}\\
\includegraphics[bb=0bp 0bp 900bp 1050bp,scale=0.25]{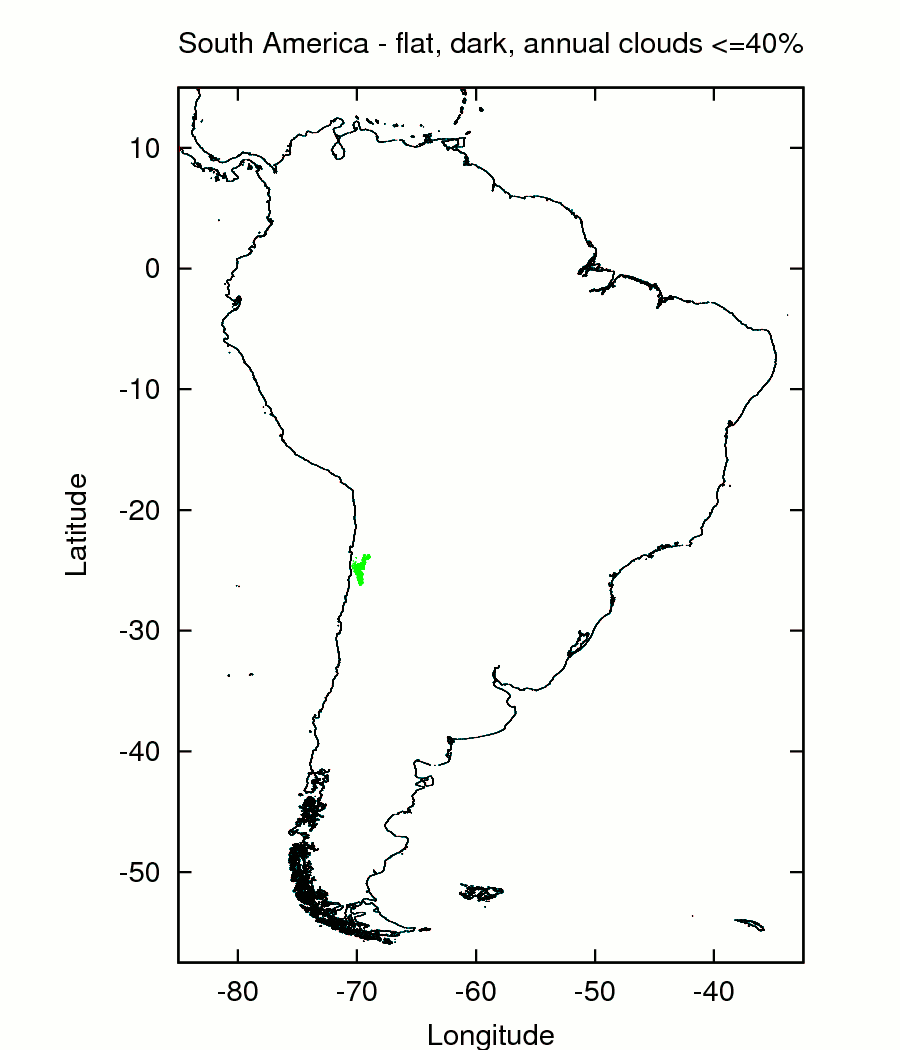}\includegraphics[bb=0bp 0bp 1100bp 1050bp,scale=0.25]{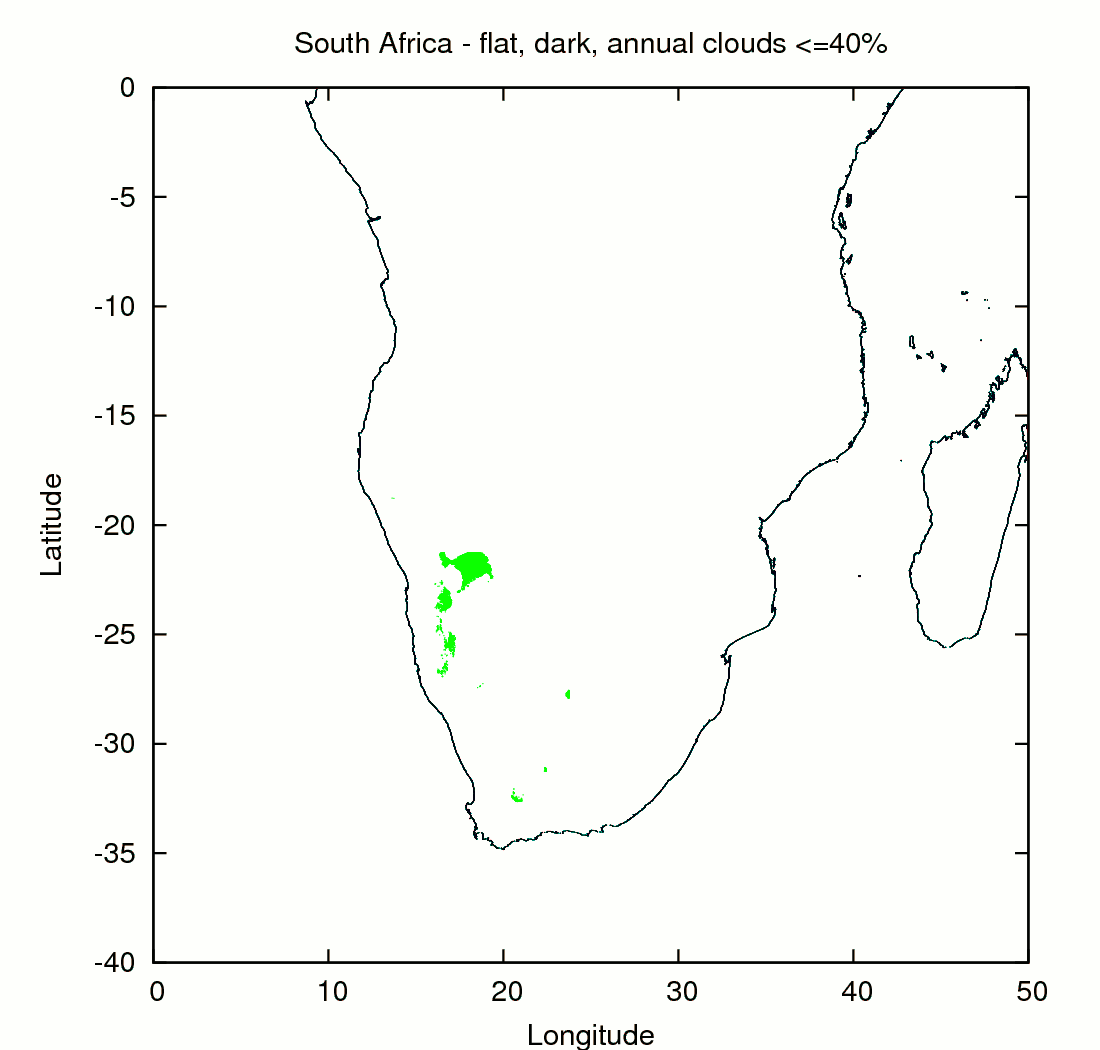}
\par\end{centering}

\caption{Regions in the southern hemisphere found by \citet{Bulik2009} that
comply to CTA's height, flatness and darkness criteria with an annual
cloud coverage of less than 35\% and 40\% respectively.\label{fig:Bulik}}

\end{figure}

CTA-south requires flat sites (less than 150~m variation) of 10~$\mbox{km}^{2}$
that have an altitude of between 1500~m and 3800~m in the southern
hemisphere with more than 70\% of nights suitable for observation
\citep{CTA2010}. The global CTA site search by \citet{Bulik2009}
only gave a number of sites along the west coast of Southern Africa
and along the west coast of Chile that meets the geographic criteria
and has an annual cloud coverage of between 30\% and 35\% and between
35\% and 40\%, as shown in Figure \vref{fig:Bulik}.

\begin{figure}
\begin{centering}
\includegraphics[height=0.8\paperheight]{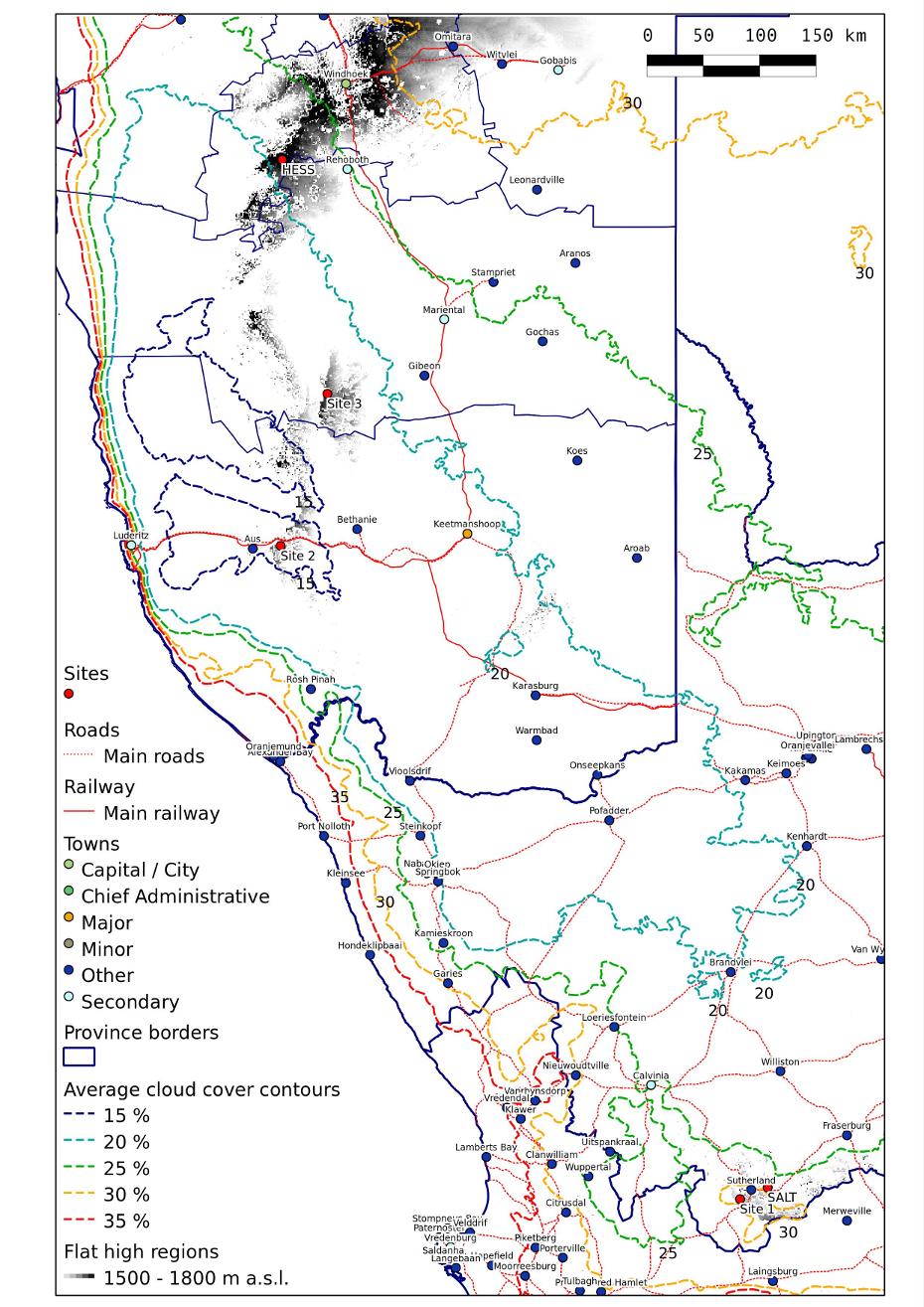}
\par\end{centering}

\centering{}\caption{Average night-time cloud coverage contour map of Southern Africa using
the MODIS satellite data. The gray-scale overlay shows regions with
an altitude between 1500~m (white) and 1800~m (black) a.s.l. and
a gradient of less than 2 degrees. Altitudes higher than 1800~m are
also black. \label{fig:Contour-map}}

\end{figure}

\begin{figure}
\includegraphics[width=1\columnwidth]{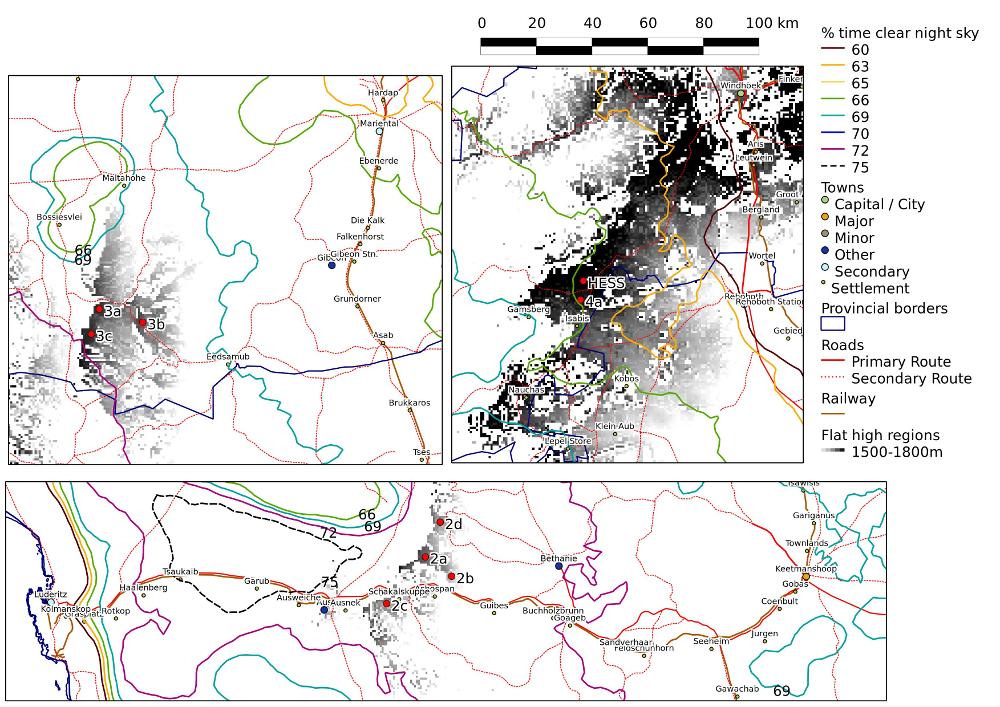}

\caption{Contour map of the percentage of the time the night sky is clear for
the three possible regions for the CTA in Namibia.\label{fig:Namibian3}}

\end{figure}

However, \citet{Bulik2009} used the annual cloud coverage compiled
by the International Satellite Cloud Climatology Project that has
a pixel size of about 2.5 by 2.5 degrees in longitude and latitude.
To obtain more detailed cloud coverage maps of Southern Africa, the
5 by 5~km cloud fraction data given by the MODIS cloud data product
for the years 2001 to 2010 are used%
\footnote{The satellite data were projected on a $90^{\prime\prime}\times90^{\prime\prime}$
gr{\small id (about $2.5\times2.5$~km) using nearest neighbour interpolation.
The sky was assumed clear if all the pixels within a radius of }$3^{\prime}$
(about 5~km) were clear. {\small The satellite data consist of only
one or two observations per night. Because there are more dark hours
in winter that in summer, each night was weighted with the length
of the night (using astronomical twilight), which was calculated using
the CBM model of \citet{Forsythe1995}. The cloud coverage of each
year from 2001 to 2010 is calculated separately. The average and standard
deviation on the average is then calculated using the 10 cloud coverage
maps.}%
}. Figure \ref{fig:Contour-map} shows the average nightly cloud coverage
along the west coast of Southern Africa. Flat regions (with a gradient
of less than two degrees to neighbouring points) that are higher than
1500~m~a.s.l. are also shown%
\footnote{{\small The same GTOPO30 data with a $30^{\prime\prime}$ (about 1~km)
resolution are used as used by \citet{Bulik2009}. }%
}. It can be seen that there are only four regions of interest: (1)
Sutherland as the only region in South Africa, (2) the region between
Luderitz and Keetmanshoop that has the fewest clouds, (3) the region
in the middle of Namibia and (4) the large region south of Windhoek
which has the highest altitudes and where H.E.S.S.~is located. The
altitude and cloud coverage of the four regions are given in Table
1 and the three Namibian regions are shown in more detail in Figure
\ref{fig:Namibian3}. 

The first two regions seem to be the best regions in South Africa
and Namibia respectively and will be discussed in more detail in the
next two chapters. 

The third region has slightly more clouds than region 2 and is slightly
higher (as shown in Figure \vref{fig:contour_3}), but is more difficult
to access (accessible from Maltahohe by the C14) and has much less
infrastructure. This site will therefore not be further considered.

\begin{figure}
\begin{minipage}[c][1\totalheight][t]{0.5\columnwidth}%
\includegraphics[width=1\columnwidth]{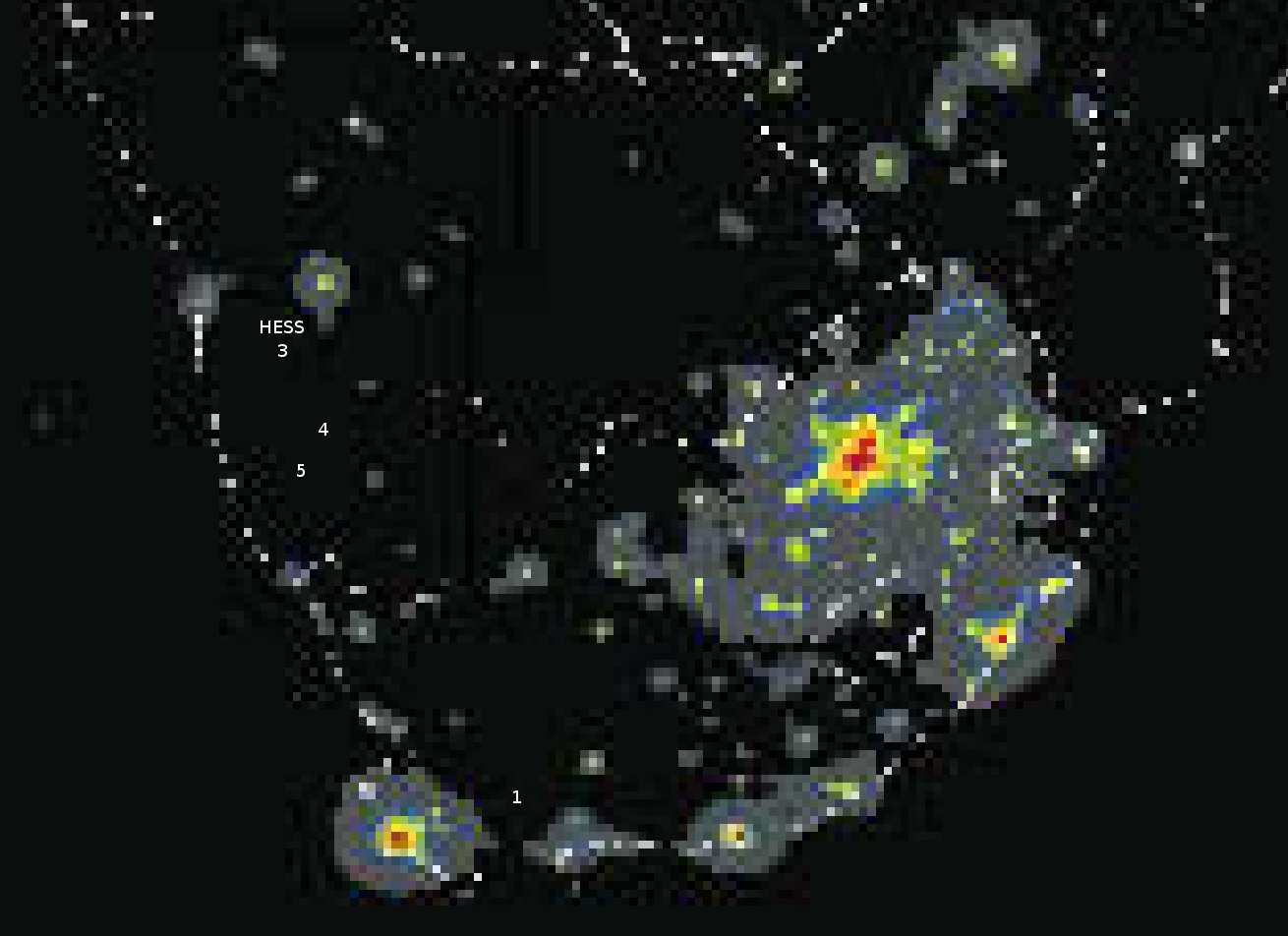}

\caption{Artificial night sky brightness at sea level for Southern Africa.
The map has been computed for the photometric astronomical V band,
at the zenith, for a clean atmosphere with an aerosol clarity coefficient
$K=1$. The calibration refers to 1996-1997. Country boundaries are
approximate \citep{Cinzano2001}.\label{fig:Artificial-night-sky}}
\end{minipage}~~~~%
\begin{minipage}[c][1\totalheight][t]{0.45\columnwidth}%
\includegraphics[bb=30bp 280bp 270bp 520bp,clip,width=1\columnwidth]{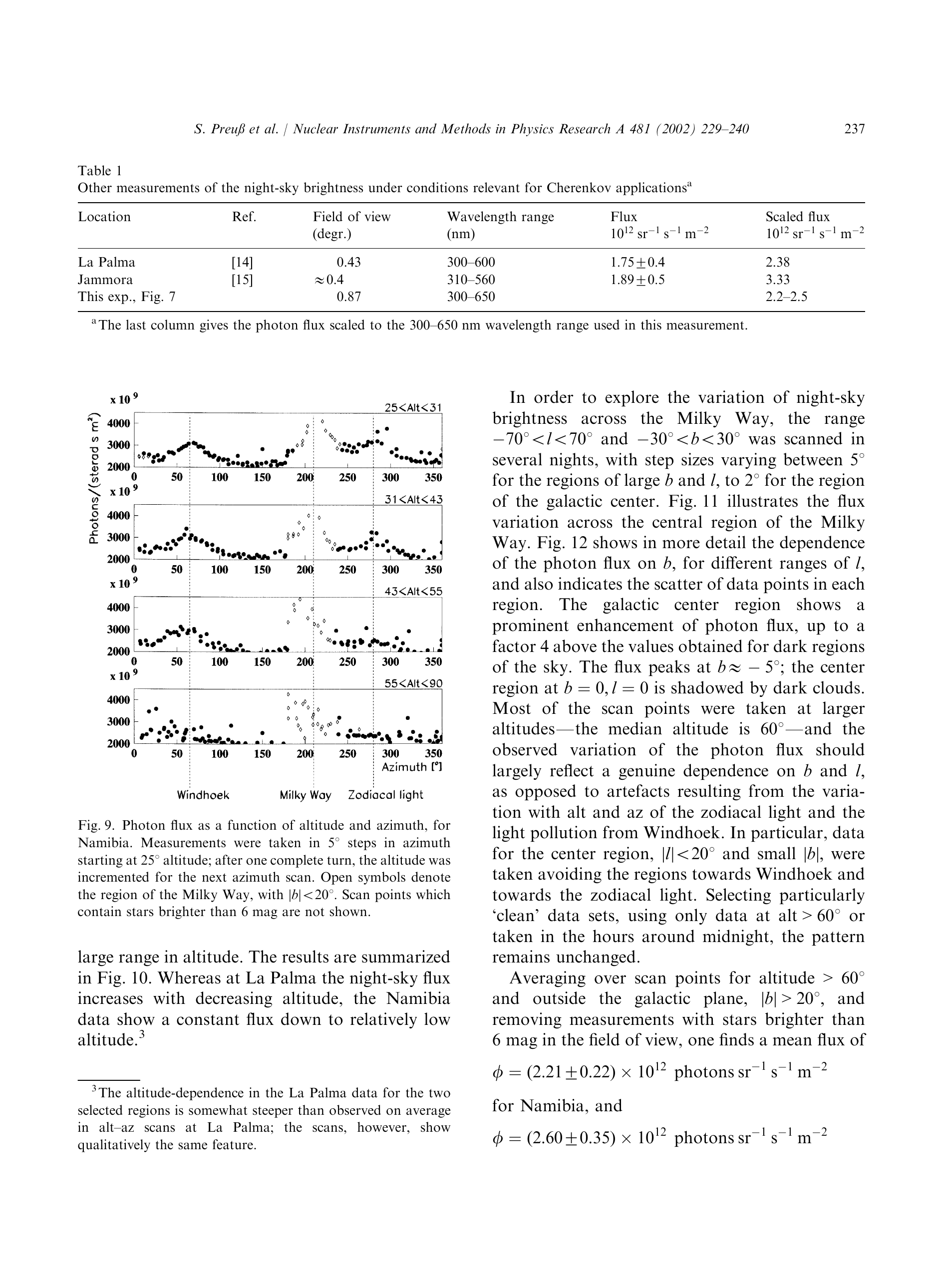}

\caption{Photon flux as a function of altitude and azimuth, for the H.E.S.S.~site
in Namibia \citep{Preuss2002}.\label{fig:Photon-flux-as}}
\end{minipage}
\end{figure}

The fourth region is in the highlands south of Windhoek where H.E.S.S.~is
located. In general, the altitude becomes lower to the south-east
and the cloud coverage becomes less to the west. The position of H.E.S.S.~seems
like a good compromise, with an altitude of 1840~m and clear skies
64\% of the time. However, the flat areas above 1800~m in the region
of H.E.S.S.~are not as flat as the other regions (see Figure \vref{fig:contour_4}).
There is a number of flat areas to the south between 1700~m and 1800~m~a.s.l.
(for example area A shown in the figure at 1720-1780~m~a.s.l.).
Because the H.E.S.S.~region is already well known, this region will
not be considered further, but will be used as a reference when considering
the first two regions.\newpage{}

\begin{minipage}[c][1\totalheight][t]{0.5\columnwidth}%

\section{Night Sky Background}

The artificial night sky brightness for almost the whole of Namibia,
as well as the Northern Cape province, is very low as can be seen
from Figure \ref{fig:Artificial-night-sky}, due to the low population
density of these regions (Figure \ref{fig:Density-of-people}). What
is also important is the distance from the site to large towns or
cities, as this influences the zenith angle at which the scattered
light from the towns will influence the night sky background. For
example, H.E.S.S.~is just over 100 km away from Windhoek and \citet{Preuss2002}
found that at the H.E.S.S.~site the night sky background light is
independent of the zenith angle, for zenith angles up to $70^{\circ}$.
However, Windhoek is still close enough to have an effect as can be
seen from Figure \ref{fig:Photon-flux-as}. The Namibian regions south
of H.E.S.S.~(regions 2 and 3) are much farther from any large town
(see Figure \ref{fig:Density-of-people}) and will therefore be even
darker at large zenith angles. Region 1 in South Africa is at about
the same distance from the light pollution of the Western Cape than
H.E.S.S is from Windhoek, and its background properties are therefore
expected to be similar, if not slightly worse, than H.E.S.S.%
\end{minipage}%
\begin{minipage}[c][1\totalheight][t]{0.5\columnwidth}%
\begin{figure}[H]
\noindent \begin{centering}
\includegraphics[bb=0bp 0bp 600bp 525bp,clip,width=1\columnwidth]{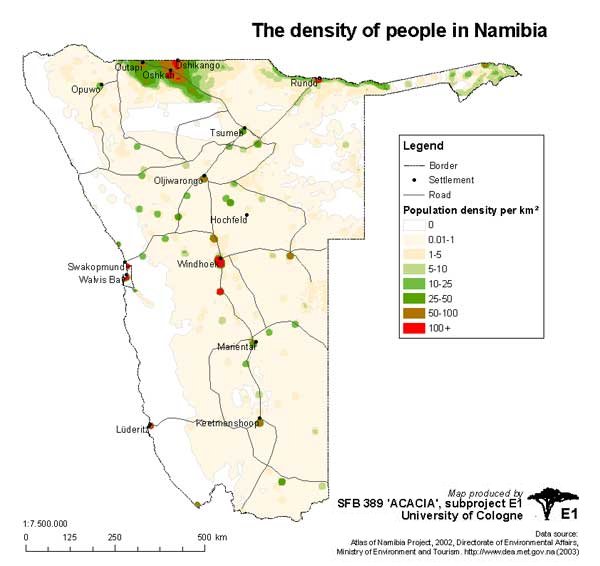}
\par\end{centering}

\noindent \begin{centering}
\caption{Population density of Namibia.\label{fig:Density-of-people}}

\par\end{centering}

\end{figure}
\end{minipage}

\section{Seismic events}

\begin{figure}[h]
\noindent \centering{}\includegraphics[width=0.85\columnwidth]{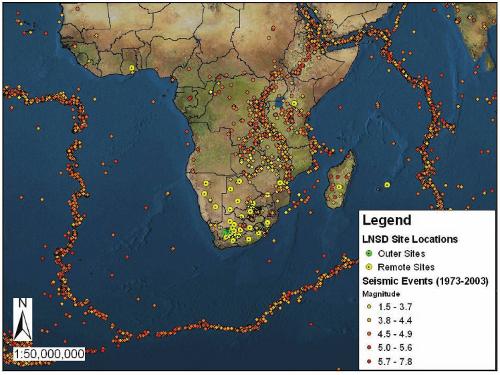}\caption{Map of seismic events close to Southern Africa from 1973 to 2003 \citep{NRF2005}.\label{fig:Map-of-seismic}}

\end{figure}

Southern Africa lies in the middle of a continental plate as shown
in Figure \ref{fig:Map-of-seismic}, resulting in the Northern Cape
and Namibia being seismically very stable. The events in the north-eastern
region of South Africa are relatively small and due to mining activities.

\newpage{}

\section{Political and economic conditions}

Both South Africa and Namibia are politically and economically stable
as can be seen from the World Bank reports (Appendix B.1 and B.2)
and the Euler Hermes country risk profiles (Appendix B.3) of the two
countries. Due to the strong South African economy, living costs in
Southern Africa are comparable, or even cheaper, than the costs in
developed countries. According to the latest cost-of-living survey
done by Mercer%
\footnote{http://www.mercer.com/costoflivingpr%
}, Windhoek is given as the tenth cheapest city out of 214 across the
world, with Johannesburg and Cape Town ranked 64th and 44th cheapest.
More detail concerning the cost of water, electricity, transportation,
fuel, telecommunication, human resources (including information on
the labour act) and taxation in Namibia for different regions is given
by the Namibia Investment Centre \citeyearpar{Centre2008}.

\begin{figure}[h]
\subfloat[All recorded offenses]{\includegraphics[width=0.32\columnwidth]{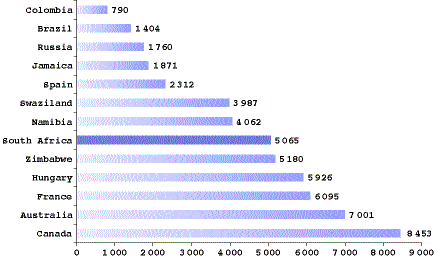}

}\subfloat[Murders ]{\includegraphics[width=0.32\columnwidth]{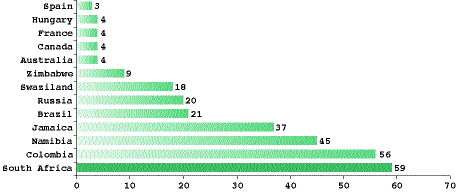}

}\subfloat[Robberies and violent thefts]{\includegraphics[width=0.32\columnwidth]{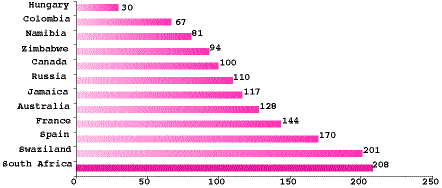}

}

\subfloat[Serious assaults]{\includegraphics[width=0.32\columnwidth]{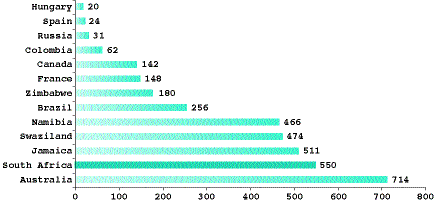}

}\subfloat[Burglaries]{\includegraphics[width=0.32\columnwidth]{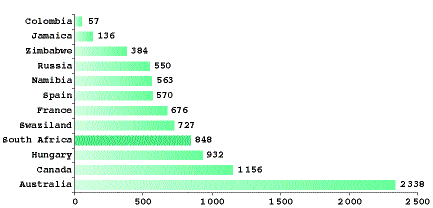}

}\caption{Number of crimes recorded per 100 000 of the population by Interpol,
1998$^{4}$.\label{fig:CrimeComparison}}

\end{figure}

Apart from \textbf{criminality}, the four possible CTA sites are relatively
safe regarding natural disasters, wild animals and regional diseases.
Based on the Interpol statistics%
\footnote{http://www.iss.co.za/pubs/CrimeIndex/01Vol5No1/World.html%
} given in Figure \vref{fig:CrimeComparison}, South Africa has high
levels of property crime and extraordinary high levels of violent
crime. Namibia has lower crime figures, but is not far behind South
Africa with regard to violent crimes. There are conflicting claims
that the crime in both countries have significantly decreased or increased
during the last decade. Crime statistics for South Africa are readily
available (e.g. at http://www.saps.gov.za), but the statistics for
Namibia is not publicly available. The latest crime statistics for
the Sutherland region (with a population of about 2000) seem to be
similar to that reported by Interpol. The Sutherland police station
reported a yearly average (for the period March 2003 to March 2010)
of 3 murders per year (143/100k), 12 cases of serious assault (560/100k)
and 41 cases of theft (2k/100k). Table \vref{tab:KarasCrime} gives
some recent crime statistics for the Karas region (site 2) as given
by the police commissioner of the Karas region, Josephat Abel%
\footnote{http://mobi.namibiansun.com/story/karas-reports-decline-crime%
}, which is also similar to that reported by Interpol. 

However, it is important to note that most of the crimes are concentrated
in economically disadvantaged areas and that, due the remoteness and
low population density of the possible CTA sites in Namibia, the crime
in these regions is much lower than the national average. 

\begin{table}[h]
\caption{2009 and 2010 crime statistics for the Karas region (site 2)$^{5}$.
The total number of cases reported as well as the cases per 100 000
people (assuming a population of 70 000) is given.\label{tab:KarasCrime}}

\noindent \centering{}\begin{tabular}{|c|c|c|}
\hline 
Year & 2009 & 2010\tabularnewline
\hline 
 & Total (per 100k)  & Total (per 100k)\tabularnewline
\hline
\hline 
Serious assault & 885 (1264) & 511 (730)\tabularnewline
\hline 
Common assault  & 776 (1109) & 420 (600)\tabularnewline
\hline 
Murders & 28 (40) & 12 (17)\tabularnewline
\hline
\end{tabular}
\end{table}

\newpage{}

\section{Astronomy in Southern Africa}

The major astronomy facilities in Southern Africa are:
\begin{description}
\item [{South~African~Astronomical~Observatory}] (SAAO) in Cape Town
and Sutherland, South Africa, is the premier optical/infrared astronomy
facility in Africa and plays a leading role in the promotion of astronomy
on the continent. SAAO promotes fundamental research in astronomy
and astrophysics at a national and international level through the
provision and utilisation of a world-class astronomical facility. 
\item [{Southern~African~Large~Telescope}] (SALT) at Sutherland, South
Africa, is the largest single telescope in the southern hemisphere
and will enable astronomers to be internationally competitive in ground-based
astronomy well into the 21st century. SALT provides a first-class
facility for fundamental research in Africa.
\item [{Karoo~Array~Telescope}] (MeerKAT) is a SKA prototype telescope
that will consist of a 64 dish radio telescope array in the Northern
Cape, with seven 12~m dishes (KAT-7) planned to be operational by
mid-2012. 
\item [{Hartebeesthoek~Radio~Astronomy~Observatory}] (HartRAO) close
to Johannesburg, South-Africa, is the only major radio observatory
in Africa (prior to MeerKAT), having a 26m-diameter centimeter-wavelength
radio telescope. The facility also supports an active program of space
geodesy, including operation of a Satellite Laser Ranger (SLR) and
a GPS network.
\item [{High~Energy~Stereoscopic~System}] (H.E.S.S.) in Namibia is the
largest and most sensitive gamma-ray telescope in the world.
\item [{Square~Kilometre~Array}] (SKA) project has short-listed South
Africa, together with Australia, as potential host countries. A final
decision in this regard is expected in 2012. This will be by far the
biggest radio telescope in the world, and in many ways the largest
terrestrial astronomical endeavour ever attempted.
\end{description}
There are only 101 people (Oct. 2010) with a PhD working in the field
of astronomy, distributed across 13 institutions. The National Research
Foundation (NRF) are therefore placing much emphasis on \textbf{human
capital development} in Astronomy:
\begin{quote}
{}``Astronomy has received high priority status in South Africa.
Major projects in support of this include Southern African Large Telescope,
H.E.S.S.~and the efforts relating to KAT and Square Kilometer Array
(SKA). The country runs the risk of not being able to fully utilise
or maximise the benefits from these facilities, owing to a lack of
adequate human capital. This risk implies that that the country may
own the hardware, but will be subject to a form of \textquotedblleft{}knowledge
colonisation\textquotedblright{} from international quarters, many
of whom already possess a critical mass of requisite skills.'' %
\footnote{http://www.nrf.ac.za/projects.php?pid=54%
}
\end{quote}
South Africa has a \textbf{DST/NRF~research~chair} (SARChl) initiative,
which is a brain-gain and research capacity-development intervention
by the Department of Science and Technology (DST) that is being administered
by the National Research Foundation (NRF). It is geared towards the
development of high-level human capital to support key areas of competency
in South Africa. Currently there are a number of chairs in astronomy,
including astrophysics and space physics, multi-wavelength extragalactic
astronomy and radio astronomy.

The \textbf{International Astronomical Union\textquoteright{}s Office
for Astronomy Development} is also hosted by South Africa.
\begin{quote}
\textquotedbl{}I am particularly pleased that our Executive Committee
chose South Africa and the South African Astronomical Observatory.
South Africa is a role model for us because it combines world-class
astronomical research facilities with a pioneering programme of astronomical
outreach.\textquotedbl{} Prof George Miley, IAU Vice President for
Development and Education.

\textquotedbl{}The National Research Foundation is pleased to host
this office, which is significant not only for South Africa and Africa
but for the entire developing world. We are pleased to be associated
with this effort to use astronomy to foster education and capacity
building globally over the next decade at least. Astronomy remains
one of the sources of inspiration for young people who take up careers
in science and technology and go on to contribute positively in society.
We are looking forward to working with the astronomy community in
developing interest in astronomy and science and technology in general\textquotedbl{}
Dr Albert van Jaarsveld, President and CEO of the National Research
Foundation.%
\footnote{http://www.saao.ac.za/no\_cache/public-info/news/news/article/185/%
}
\end{quote}

\subsection{South African Government's commitment to astronomy}

In the above-mentioned projects, South Africa has invested heavily
in astronomy (compared to the size of the economy). 
\begin{quote}
\textquotedblleft{}We chose to invest heavily in science and astronomy,
because of its role in development, not only within South Africa,
but all across Africa. Big astronomy projects such as SALT, MeerKAT
and SKA entail major capacity development programmes in order to train
the next generation of engineers and astronomers from all over Africa.
In South Africa, people in the astronomy field, from those working
on the ground to the highest levels of government, share the vision
that astronomy will play a significant role in the development of
society.\textquotedblright{} (South African Science and Technology
Minister, Naledi Pandor, at the launch of the International Astronomical
Union Global Office of Astronomy for Development at the headquarters
of the SAAO in Cape Town this year.)
\end{quote}
The Department for Science and Technology has launched the Astronomy
Geographic Advantage Programme (AGAP) with the aim to use the excellent
viewing conditions on the sub-continent and the depth of engineering
and scientific talent available here to attract international astronomy
projects to our region. Major successes thus far have been the H.E.S.S.~telescope
in Namibia and SALT in Sutherland. The SKA project, if successful
would be the jewel in the crown of AGAP.

\subsubsection{Astronomy Geographic Advantage Bill}

{}``The {[}Astronomy Geographic Advantage Bill{]} gives the Minister
of Science and Technology the power to declare astronomy advantage
areas in order to ensure that large-scale and globally important astronomy
facilities are protected from developments that might interfere with
their research activities.

The benefits of the legislation include the protection of large-scale
investments already made in astronomy; the preservation of an environment
for a global astronomy hub that will continue to attract international
investment; and the provision of a competitive advantage to help South
Africa become the preferred host of the full Square Kilometre Array
(SKA) in the Northern Cape province, also home to the Southern African
Large Telescope (SALT), the largest single optical telescope in the
Southern Hemisphere. 

To do this, the Act provides for:
\begin{itemize}
\item Developing the skills, capabilities and expertise of those engaged
in astronomy and related scientific endeavours in Southern Africa. 
\item Identifying and protecting areas in which astronomy projects of national
strategic importance can be undertaken. 
\item The declaration and management of astronomy advantage areas. 
\item Providing a framework for the establishment of a national system of
astronomy advantage areas to ensure that geographic areas highly suitable
for astronomy and related scientific endeavours owing, for example,
to their high atmospheric transparency, low levels of light pollution,
low population density or minimal radio frequency interference are
protected, preserved and properly maintained. 
\item Enhancing the country's geographic advantage by restricting activities
that cause or could cause light pollution or radio frequency interference,
or might interfere in any other way with astronomy and related scientific
endeavours in these areas. {}`` (Statement issued by the Department
of Science and Technology, June 2008%
\footnote{http://www.info.gov.za/speeches/2008/08062609451003.htm%
})
\end{itemize}

\subsection{Multi-wavelength astronomy}

With the planned commissioning of MeerKAT in 2013, Southern Africa
will host a world-class radio (MeerKAT), optical (SALT) and gamma-ray
(H.E.S.S.) telescope. This opens a unique opportunity to simultaneously
observe variable sources across the three wavelengths, making true
multi-wavelength astronomy possible. Below are a few examples of the
study of astrophysical sources within a multiwavelength context given
by \citet{Venter2011}. These example shows how multiwavelength observations
result in a complementary approach for source discovery, as well as
to disentangle details of particle injection, acceleration, transport,
and radiation within these sources.

\subsubsection{Pulsars}

The high-energy (HE) astrophysics domain was revolutionized by the
launch of \emph{Fermi} Large Area Telescope (LAT) on 11~June 2008.
The second \emph{Fermi} Catalog is in production and contains nearly
1,900 gamma-ray sources. Among the numerous discoveries are almost
100 gamma-ray pulsars. Almost a third of these have been found using
blind periodicity searches in the gamma-ray data at \emph{a priori}
source positions. Follow-up observations in the radio found pulsations
from only three of these, and radio upper limits for 23 more. This
means that the bulk of these gamma-ray-selected pulsars is still radio-faint,
either due to beaming effects, or intrinsically weak radio emission
properties. Deep searches with (future) radio telescopes such as MeerKAT
and SKA in the \emph{Fermi} LAT radio-faint pulsar error boxes may
reveal weak radio emission and thus discover more radio pulsars. This
will impact pulsar population modelling, since one important prediction
of such models is the fraction of radio-quiet (Geminga-like) pulsars.

A number of millisecond pulsars (MSPs) have now been found by observing
non-variable, unassociated \emph{Fermi} sources at high latitudes
with radio telescopes and searching for pulsed radio signals. This
is in contrast to the traditional method where gamma-ray data in the
direction of known radio pulsars are searched using the radio ephemeris.
These discoveries suggest a new mode of discovery, and future radio
follow-up of \emph{Fermi} unidentified bright sources with pulsar-like
spectra may lead to many more (millisecond) pulsar discoveries.

The detection of many gamma-ray pulsars for the first time affords
the opportunity for true multiwavelength modelling. One example is
being able to constrain the pulsar geometry (inclination and observer
angles $\alpha$ and $\zeta$) using complementary approaches: radio
light curve as well as polarization swing data and application of
the rotating vector model provide constraints on $\alpha$ and $\beta=\zeta-\alpha$,
while geometric gamma-ray and radio light curve modelling provide
additional constraints on $\alpha$ and $\zeta$. Further constraints
on $\zeta$ may be derived using X-ray PWN data.

Previous predictions of low inverse Compton VHE components for certain
pulsar parameters, as well as HE spectral fits with cutoffs around
a few GeV for all of the current \emph{Fermi} pulsars impacted negatively
on expectations for VHE emission from pulsars. However, a very recent
detection of the Crab pulsar at several hundred GeV by the VERITAS
Collaboration now suggests that pulsar science may indeed be possible
using ground-based Cherenkov telescopes. The lower energy threshold
of CTA will greatly aid in constraining the spectral shapes and therefore
magnetospheric properties such as the electric field and radius of
curvature of particle orbits.

\subsubsection{Active Galactic Nuclei}

The active galaxy PKS~2155-304 was among the first sources detected
by H.E.S.S., and an extreme gamma-ray outburst at 10 times the usual
flux level was reported subsequently. This BL Lac object has furthermore
been observed in follow-up multiwavelength campaigns. The first of
these involved H.E.S.S., the \emph{Rossi X-ray Timing Explorer} (RXTE)
satellite, the \emph{Robotic Optical Transient Search Experiment}
(ROTSE), and the \emph{Nan\c{c}ay Decimetric Radio Telescope} (NRT).
During this campaign, the object appeared to be in a quiescent state,
exhibiting flux variability in all wavebands, flux-dependent spectral
changes in the X-rays, and also a transient X-ray event with a 1,500-second
timescale. During a second campaign, \emph{Fermi} LAT provided MeV
to GeV coverage, while the \emph{ATOM} telescope and RXTE and \emph{Swift}
observatories provided concurrent optical and X-ray observations.
This time, the object was in a low X-ray and VHE state, but the optical
flux was relatively high. There was a clear correlation of optical
and VHE fluxes (hinting at a common parent particle population), and
a possible correlation between the X-ray flux and HE spectral index,
but none between the X-ray and VHE components (in contrast to what
is seen in many other blazars). 

This example indicates how vital simultaneous observations in various
energy regimes are to help constrain the source properties. In particular,
an optical telescope such as SALT, with its superior sensitivity,
allows probing of polarization variations with time. This will help
to distinguish between multiple active emission sites, and constrain
their structure, location, geometries, environments, evolution, local
magnetic field properties, and the particle acceleration and cooling
processes, as well as guide construction of future multizone models.
In addition, SALT spectroscopy and images will allow identification
of new high-redshift AGN associated with high-latitude unidentified,
variable \emph{Fermi} sources.

\subsubsection{Astrophysical Lepton and Hadron Colliders}

SNRs have long been thought to be accelerators of hadronic CRs. However,
electrons are also believed to be accelerated to very high energies
in these systems. Gamma-ray signals from such SNRs may therefore originate
from either hadronic or leptonic CRs interacting with ambient matter,
soft radiation fields, or electromagnetic fields -- or a combination
of both -- and it is difficult to discriminate between the two scenarios.
One example is the H.E.S.S.~detection of VHE flux in the direction
of the old W28 SNR, known to be interacting with surrounding molecular
clouds along its north and northeastern boundaries. The gamma-ray
flux revealed a partial correlation with the radio emission, but a
good correlation with NANTEN $^{12}\mbox{CO(}J=1-0$) data. The gamma-ray
flux therefore follows the molecular gas structure, probably pointing
to interaction of hadronic CRs with the target material in this case.

\emph{Fermi} has also begun to contribute to the debate of the relative
contribution of hadronic vs. leptonic parent populations to the observed
gamma-ray signals from several SNRs. However, \emph{Fermi} source
identification and characterization suffer significantly from contributions
of the diffuse, structured Galactic gamma-ray background that traces
molecular clouds near the Galactic Plane, while this diffuse background
is very faint at TeV (H.E.S.S.~and CTA) energies. MeerKAT's survey
potential will discover new resolved sources of energetic particles
(e.g. SNRs and colliding winds), so that a combination of MeerKAT,
H.E.S.S.~/ CTA, and CO observations should resolve many more hadron
colliders in the Galactic Plane, thereby constraining the origin of
VHE signals from several more sources. 

\theendnotes

\chapter{Sutherland, Northern Cape, South Africa}

\section{Geography}

\begin{figure}
\begin{centering}
\includegraphics[bb=0bp 0bp 894bp 550bp,clip,width=1\columnwidth]{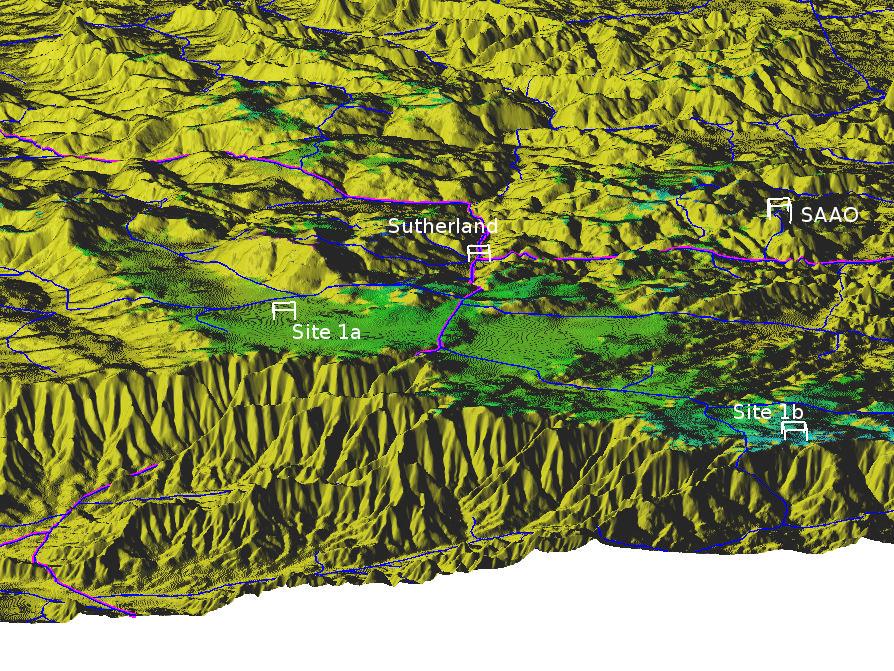}
\par\end{centering}

\centering{}\caption{Height profile of Site 1. Flat areas are colored green.\label{fig:Height-sutherland}}

\end{figure}
\begin{figure}
\noindent \raggedright{}~~%
\begin{minipage}[c][1\totalheight][t]{0.48\columnwidth}%
\noindent \begin{flushleft}
\includegraphics[bb=0bp 0bp 894bp 550bp,clip,width=1.1\columnwidth]{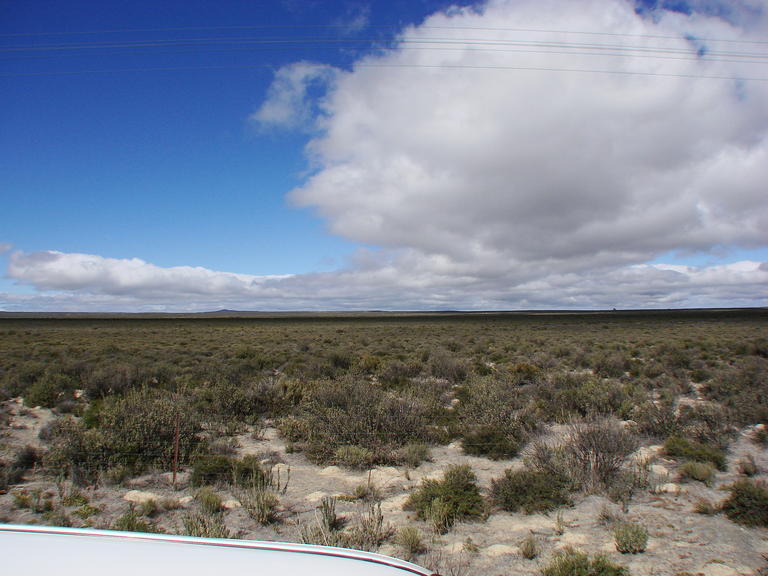}
\par\end{flushleft}

\begin{center}
\caption{Picture of site 1 taken from the road on the north.\label{fig:Picture-site1} }

\par\end{center}%
\end{minipage}
\end{figure}

South of the town Sutherland, just above the escarpment, is a large
flat region as shown in Figure \ref{fig:Height-sutherland}. The region
is about 1660~m a.s.l. in the east (1b) and drops gradually to about
1500~m a.s.l. in the west, with the large flat region (1a) just west
of the main road at about 1560m. Note that the western side is more
flat that the eastern side. Figure \ref{fig:Picture-site1} is a photo
of region (1a).

\section{Cloud coverage}

\begin{figure}
\centering{}\includegraphics[width=1\columnwidth]{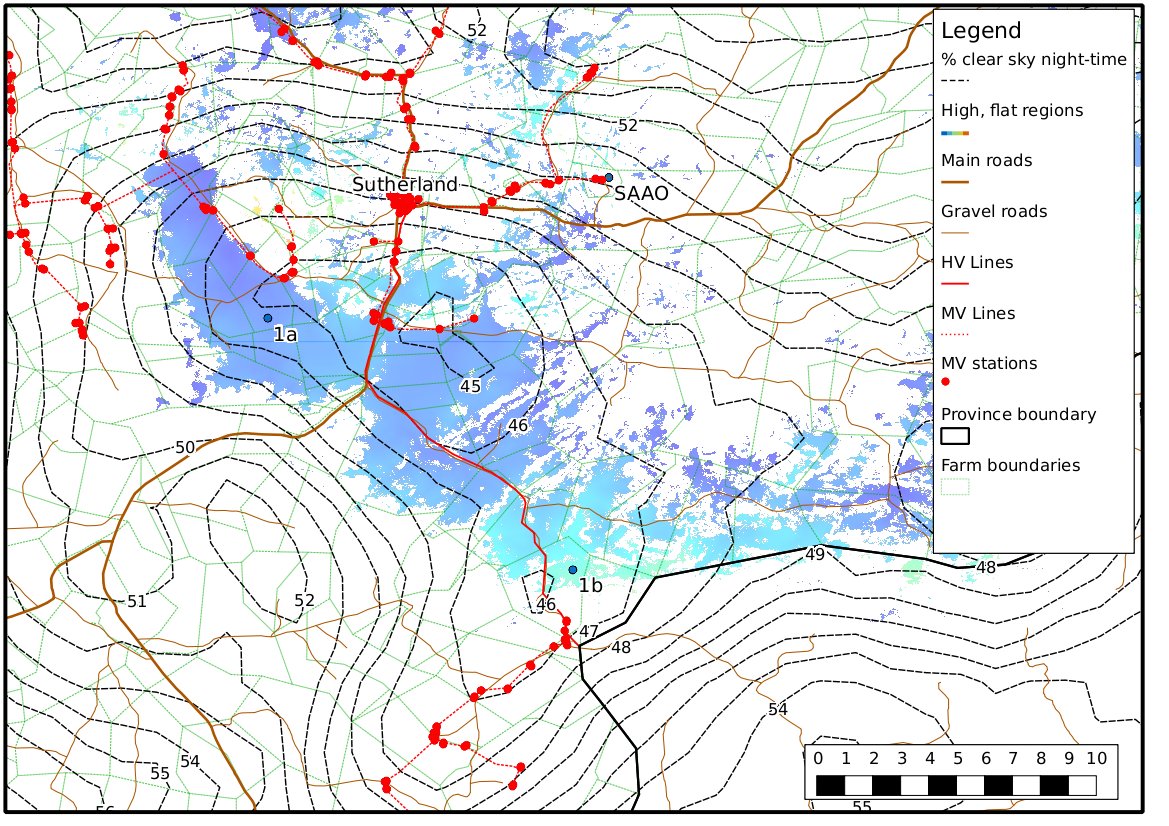}\caption{Contour map of clear sky fraction of the Sutherland region. \label{fig:Cloud1}}

\end{figure}
Figure \ref{fig:Cloud1} shows the cloud coverage contours of the
region. It can be seen that the flat region just above the escarpment
is actually more cloudy than the surrounding area, such as SAAO where
the SALT optical telescope is located.

\begin{table*}
\caption{Comparison of cloud coverage of Sutherland, SAAO and Site 1a and 1b
using satellite data and manual observations.\label{tab:Comparison-of-cloud}}

\begin{tabular}{|c|c|c|c||c|c||c||c|}
\hline 
 & \multicolumn{3}{c|}{Sutherland} & Site 1a & \multicolumn{1}{c|}{Site 1b} & \multicolumn{2}{c|}{SAAO}\tabularnewline
\hline 
 & Obs: 6h & Obs 18h & \multicolumn{4}{c||}{Satellite} & Logs\tabularnewline
\hline
\hline 
\% Cloud cover & 31.1$\pm$1.4 & 29.4$\pm$1.3 & 30.2$\pm1.1$ & 32.4$\pm1.1$ & 32.9\textbf{$\pm0.9$} & 29.0$\pm1.1$ & \tabularnewline
\hline 
\% Time clear sky & 47.7$\pm$2.9 & 43.7$\pm$2.5 & 48.3$\pm1.1$ & 46.5$\pm1.1$ & 46.0$\pm1.3$ & 49.9$\pm1.0$ & 52.3$\pm$1.0\tabularnewline
\hline
\end{tabular}
\end{table*}
In order to obtain satellite cloud coverage maps, a lot of data processing
must be done and a number of assumptions must be made. This sometimes
leads to large systematic differences between different satellite
cloud coverage maps. It is therefore important to compare satellite
data with ground-based observations. To do this, the manual day-time
observations of the Sutherland weather station%
\footnote{{\small The SYNOP (surface synoptic observations) data of the Sutherland
weather station at 8:00 and 20:00 for each day from 2001 to 2010 were
used. Each record reports the cloud coverage in eighths. }%
} and the telescope logs of three optical telescopes at SAAO were used%
\footnote{{\small The photometric hours reported by the observation logs filled
in by the users of the 20'', 30'' and 40'' telescope at SAAO were
used for the years 2005 to 2010. However, not all the telescopes were
used every night, usually due to bad weather or technical problems.
Different observers also estimated the number of photometric hours
differently. To estimate the number of photometric hours for each
night, the average was taken of the reported hours of the telescopes
that was used. Therefore, if a telescope did not report photometric
hours, but the others did, it was excluded from the average. However,
when none of the telescopes reported photometric hours, it was assumed
that there were no photometric hours that night. When there was a
large difference between the reported hours, the night was flagged
and the difference investigated and corrected for when possible. However,
it was found that these differences contributed less than 1\% to the
yearly photometric hours. Some nights reported more photometric hours
than the number of hours between astronomical twilight. In order to
compare the hours with that of \citet{Harding1974}, the calculated
photometric hours were clipped to the number of available of hours,
whenever is was larger. This however, slightly biases the results
toward a larger percentage of photometric time.}%
}. The satellite and ground-based observational cloud coverage is given
in Table \ref{tab:Comparison-of-cloud}, showing that the satellite
data do not seem to have large systematic errors. The 52\% clear time
for SAAO also corresponds to the 51\% reported by \citet{Harding1974}.
Note that the cloud coverage of Sutherland was underestimated \citeauthor{Bulik2009}'s
search \citeyearpar{Bulik2009}, with the non-clear time of 50\% being
much larger than the range 35-40\%. The reason for this is that the
much less cloudy Northern Cape to the north was on the same grid point.

\section{Weather conditions}

\begin{figure}
\begin{minipage}[c][1\totalheight][t]{0.5\columnwidth}%
\includegraphics[width=1\columnwidth]{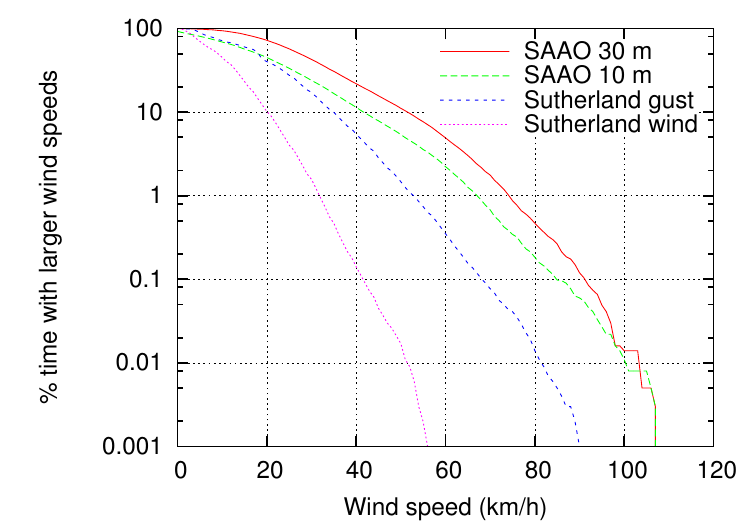}

\caption{Graph of percentage of time that wind speeds are larger than a certain
speed as recorded by the Sutherland and SAAO weather stations.\label{fig:Suth-wind}}
\end{minipage}~~~~%
\begin{minipage}[c][1\totalheight][t]{0.5\columnwidth}%
\includegraphics[width=1\columnwidth]{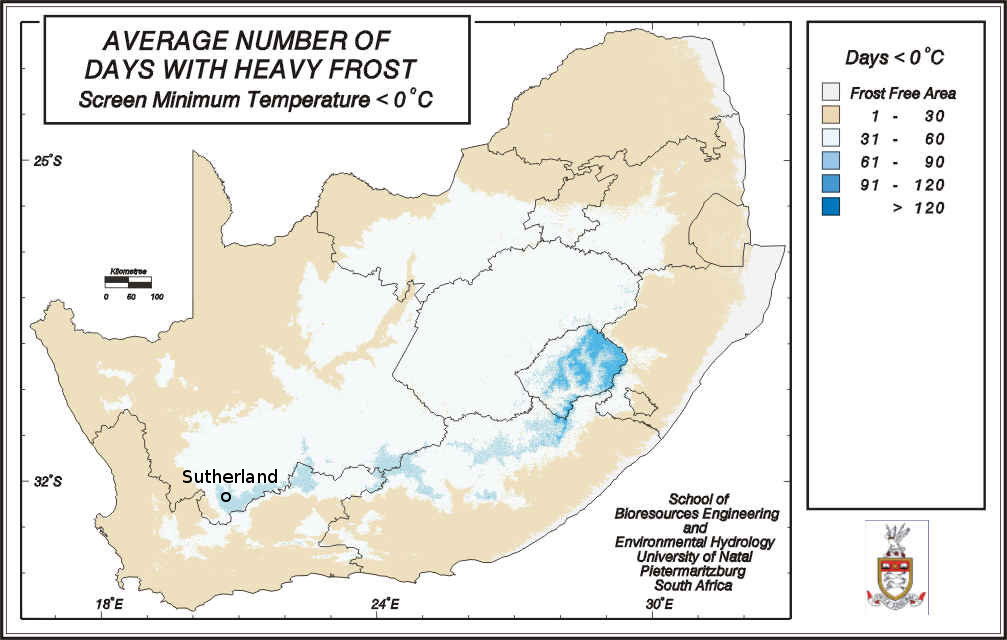}

\caption{Map of number of days with heavy frost in South Africa. \label{fig:RSA_frost}}
\end{minipage}
\end{figure}

A lot of weather data are available for the Sutherland region, due
to the weather stations at Sutherland and SAAO%
\footnote{The Sutherland weather station's five minute interval average and
gust wind speeds from Nov 2006 to Mar. 2011 and SAAO's ten minute
interval wind speeds at 10m and 30m from Apr 2008 to Apr. 2009 were
used.%
}. Figure \ref{fig:Suth-wind} shows that all the recorded wind speeds
were less than 110~km/h. Sutherland is shielded by the surrounding
hills (see Figure \ref{fig:Height-sutherland}) resulting in lower
wind speeds there than at the SAAO, which is on top of a hill. The
wind speed measured on the 30~m mast at SAAO is also larger than
on the 10~m mast, but is still less than 50~km/h for 89\% of the
time. It is expected that the wind at the proposed region will be
similar to that at the SAAO.

\begin{figure}
\begin{minipage}[c][1\totalheight][t]{0.5\columnwidth}%
\includegraphics[width=1\columnwidth]{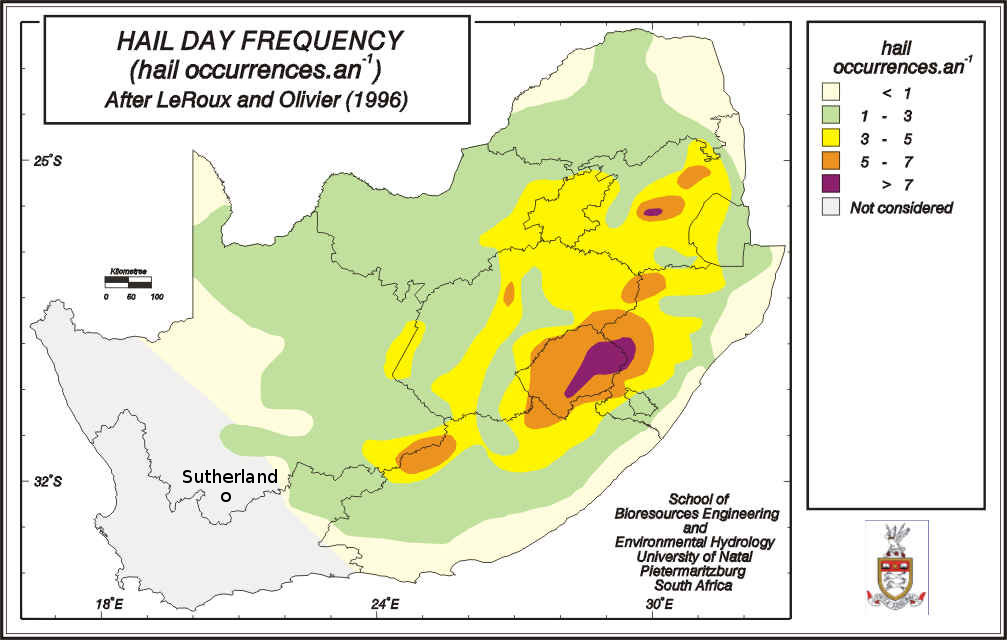}

\caption{Map of days of hail per year in South Africa. \label{fig:RSA_hail}}
\end{minipage}~~~~%
\begin{minipage}[c][1\totalheight][t]{0.5\columnwidth}%
\includegraphics[width=1\columnwidth]{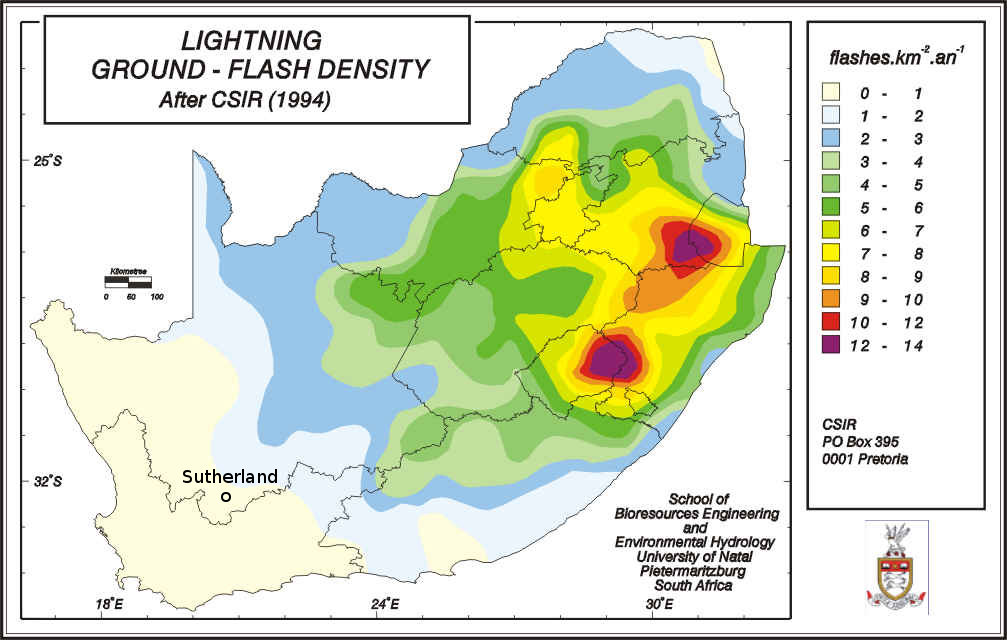}

\caption{Map of lightning ground-flash density of South Africa. \label{fig:RSA_lightning}}
\end{minipage}
\end{figure}

Figure \ref{fig:Suth_Temp} shows the average of the daily minimum
and maximum temperatures for each month at Sutherland. Because the
daily minimum temperature of Sutherland is normally below freezing
point in the winter, Sutherland receives a large amount of heavy frost
per year (Figure \ref{fig:RSA_frost}). The probability of natural
hazards is very low, as it lies outside the hail and thunder storm
region of South Africa (see Figures \ref{fig:RSA_hail} and \ref{fig:RSA_lightning}).


\section{Accessibility \& Infrastructure}

\begin{figure}
\begin{minipage}[c][1\totalheight][t]{0.5\columnwidth}%
\includegraphics[width=1\columnwidth]{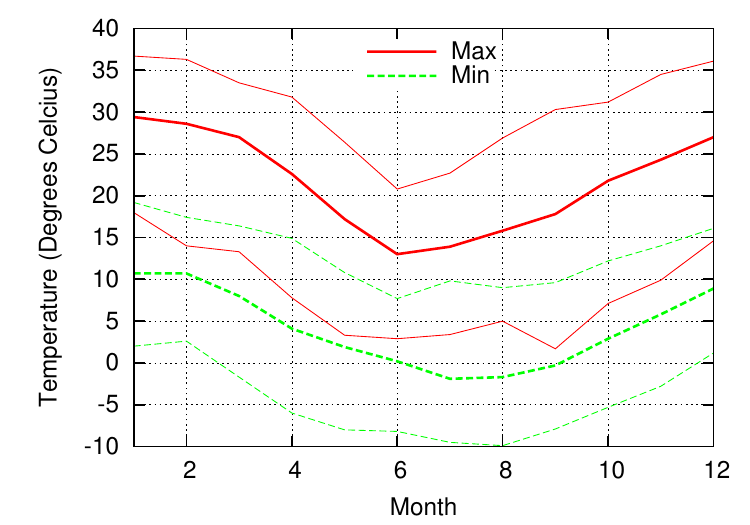}

\caption{Monthly averages and extremes of daily minimum and maximum temperatures
at Sutherland between Apr 2009 and May 2011.\label{fig:Suth_Temp}}
\end{minipage}~~%
\begin{minipage}[c][1\totalheight][t]{0.45\columnwidth}%
\includegraphics[width=1\columnwidth]{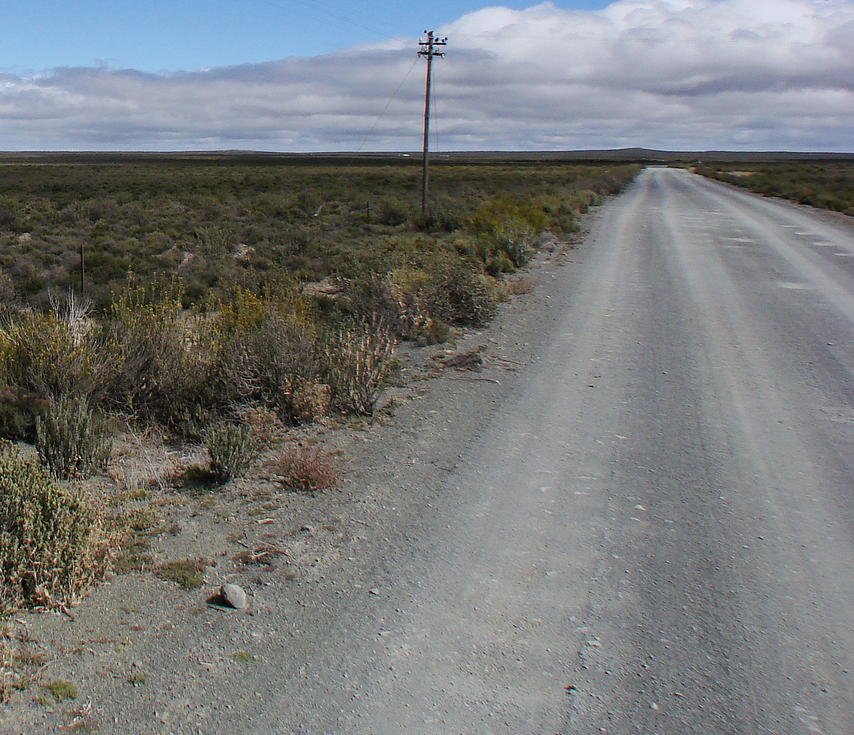}

\caption{Gravel roads around Sutherland\label{fig:Site-1:-Gravel}}
\end{minipage}
\end{figure}

The proposed CTA site is about 10~km from Sutherland along the R354.
There are good gravel roads (Figure \ref{fig:Site-1:-Gravel}) serving
the farms in the region (Figure \ref{fig:Cloud1}) on both sides of
the R354 tar road, with an air strip just east of the R354. Although
Sutherland is small (population about 2000), it offers plenty \textbf{accommodation},
has a \textbf{gasoline station, food stores} and a medical clinic.

The R354 also links the site to the N1 national road to Worcester
(220~km away,  population above 100 000) and Cape Town (330~km away).
It is about 3.5~hours drive from Sutherland to Cape Town. The nearest
\textbf{harbour} and \textbf{international airport} are at Cape Town
with a \textbf{railway} to Matjiesfontein (100~km away). The nearest
government \textbf{hospital} and \textbf{fire station} are at Calvinia
(170~km away, population of about 8 000), although Worcester also
offers a private hospital. For comparison, H.E.S.S.~is about 100~km
away from Windhoek (which is the closest large town or city) on a
gravel road.

ER24 and Netcare are two private national \textbf{emergency medical
care services} that serve the whole of South Africa. Although Sutherland
does not have an ambulance, Aerocare's air ambulance can (and has)
made emergency flights (from Bloemfontein) to the Sutherland airstrip
to transfer patients to Cape Town.

The site can be easily connected to the national \textbf{electricity
grid}, as there are various substations around site 1 (Figure \ref{fig:Cloud1})
which are at most a few kilometers away. Safe \textbf{drinking water}
is available from boreholes.

SANREN (South African National Research and Education Network) is
currently working on linking SAAO to the South African national backbone
(with speeds between 1 and 10 Gb/sec), making it easy for the site
to have \textbf{broadband internet connection}. 


\theendnotes

\chapter{Kuibis region, Karas, Namibia}

\section{Cloud coverage}

\begin{table*}
\caption{Comparison of cloud coverage for Windhoek, Keetmanshoop and Luderitz
airport using satellite data and manual observations. The sky is defined
as almost clear when one eigth of less of cloud cover is observed
or when a 5x5~km cloudless region is measurent by satellite. Observation
times are given at GMT.\label{tab:Comparison-of-cloud-1}}

\begin{tabular}{|>{\centering}p{0.15\columnwidth}|c|c||c|c|c||c|c|c|}
\hline 
 & \multicolumn{2}{c||}{Windhoek} & \multicolumn{3}{c||}{Keetmanshoop%
\footnote{Data of the all the years with more than 200 observations were used:
For 6:00 and 18:00 the years 2001, 2002, 2003, 2005, 2006, 2008, 2009
and the years 2001, 2002, 2005, 2006, 2008 were used respectively.%
}} & \multicolumn{3}{c|}{Luderitz%
\footnote{Data of the years 2000, 2001 and 2002 were used.%
}}\tabularnewline
\hline
\hline 
 & Obs: 6h & Sat: Night & Obs: 6h & Obs 18h & Sat: Night & Obs: 6h & Obs 18h & Sat: Night\tabularnewline
\hline 
\% Cloud cover & 23.3$\pm1.1$ & 26.8$\pm1.3$ & 18.6$\pm0.5$ & 19.8$\pm0.8$ & 19.5$\pm0.8$ & 28.9$\pm1.3$ & 15.8$\pm1.4$ & 20.7$\pm0.8$\tabularnewline
\hline 
\% Time clear sky & 43.2$\pm1.2$ & 56.4$\pm1.5$ & 40.8$\pm1.7$ & 48.1$\pm2.7$ & 69.7$\pm0.9$ & 49.8$\pm1.0$ & 69.2$\pm2.3$ & 45.7$\pm1.0$\tabularnewline
\hline 
\% Time almost clear sky  & 63.4$\pm1.6$ & 63.3$\pm1.4$ & 69.3$\pm1.0$ & 67.7$\pm1.5$ & 73.9$\pm0.9$ & 59.9$\pm1.1$ & 76.7$\pm2.5$ & 68.0$\pm1.1$\tabularnewline
\hline
\end{tabular}
\end{table*}

Figure \ref{fig:Namibian3} shows the three regions of interest in
Namibia. The region next to the B2 national road between Luderitz
and Keetmanshoop, about 30~km east of the town Aus, has the least
clouds of all the possible regions, with about 73\% of the night-time
cloudless. This is 57\% more time than at the Sutherland site and
13\% more time than at the H.E.S.S.~site. Assuming the moon to be
below the horizon 50\% of the time, this gives about 1200 hours per
year suitable for observation. Note that the cloud coverage for this
site was overestimated by \citeauthor{Bulik2009}'s search \citeyearpar{Bulik2009}.
This may again be due to the large grid used.

\section{Geography}

As marked in the bottom of Figure \ref{fig:Namibian3}, this region
has roughly four large flat areas. Contour maps of these areas are
given in Appendix A (Figures \ref{fig:contour_2a}-\ref{fig:contour_2d}).
Area 2a is the highest (between 1650 and 1720~m~a.s.l) but the most
difficult to access, as there seems to be no road to B2, but only
one northwards to Bethanie. Area 2b, on the other hand, is very easy
to access as it is right next to the B4. Figure \ref{fig:Photo-Site2}
shows a photo of the site as taken from the B2 road. However, it is
also lower at 1510-1590~m~a.s.l. The third area is the southern
area south of the B4 at between 1600 and 1680~m~a.s.l. The fourth
area is the most northern area at between 1590 and 1660 m~.a.s.l.
and next to the gravel road to Bethanie. 

The best area seems to be the third one, as it is close to the B2
and relatively high.

\section{Weather conditions}

\begin{sidewaysfigure}
\begin{centering}
\begin{minipage}[c][1\totalheight][t]{0.3\columnwidth}%
\includegraphics[width=1\columnwidth]{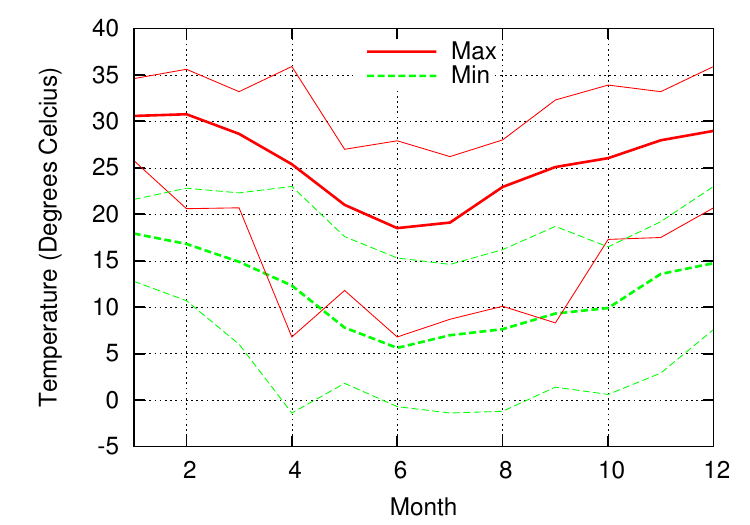}

\caption{Monthly averages and extremes of daily minimum and maximum temperatures
at Klein Aus Vista between Apr 2009 and May 2011.\label{fig:Aus_Temp}}
\end{minipage}~~~~~~~~~~%
\begin{minipage}[c][1\totalheight][t]{0.3\columnwidth}%
\includegraphics[width=1\columnwidth]{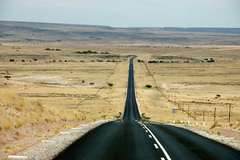}

\caption{Picture of the B4 road close to the proposed site.\label{fig:B4}}
\end{minipage}\vspace*{1cm}

\par\end{centering}

\centering{}%
\begin{minipage}[t]{0.32\columnwidth}%
\includegraphics[bb=0bp 0bp 600bp 525bp,clip,width=1\columnwidth]{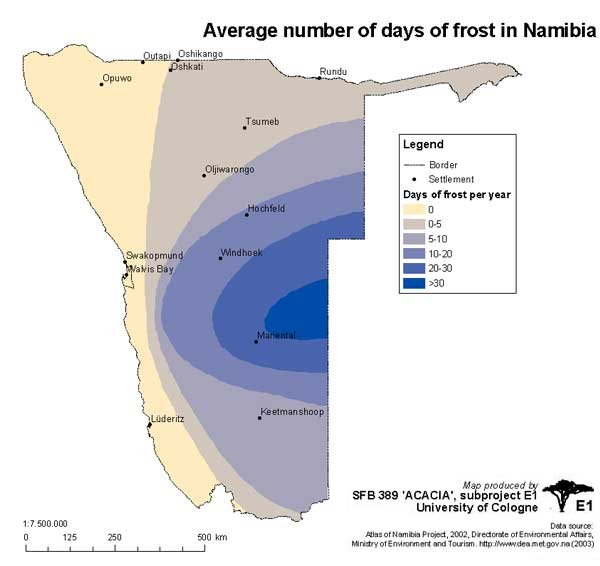}

\caption{Map of average number of days of frost in Namibia.\label{fig:Days-of-frost}}
\end{minipage}~~~~%
\begin{minipage}[t]{0.32\columnwidth}%
\includegraphics[bb=0bp 0bp 600bp 525bp,clip,width=1\columnwidth]{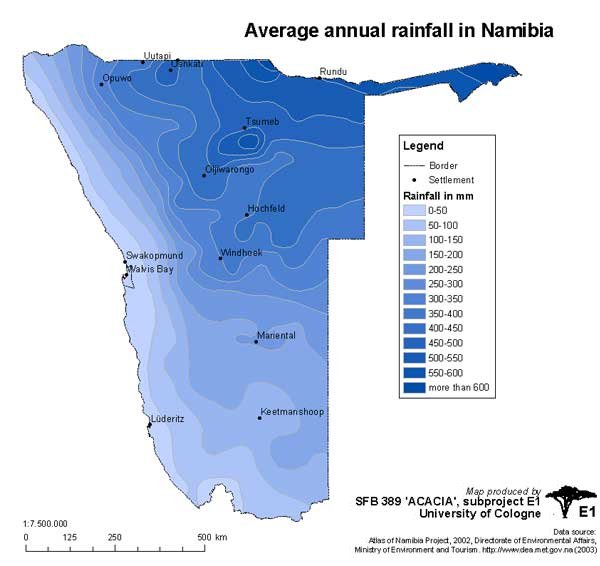}

\caption{Map of average annual rainfall in Namibia\label{fig:Rain}}
\end{minipage}~~~~%
\begin{minipage}[t]{0.32\columnwidth}%
\includegraphics[bb=0bp 0bp 600bp 525bp,clip,width=1\columnwidth]{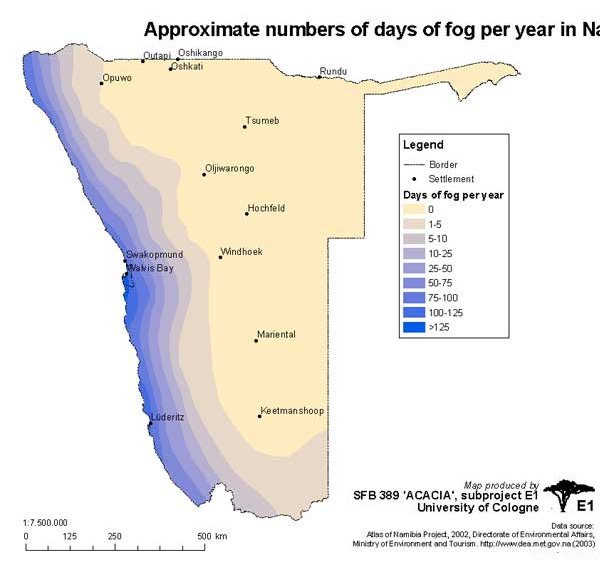}

\caption{Map of approximate number of days of fog per year in Namibia.\label{fig:Days-of-fog}}
\end{minipage}
\end{sidewaysfigure}

The Klein Aus Vista weather station is the nearest weather station
to the site (3 km west of Aus on the B4, 66 km west of the site).
From the 583 daily reports%
\footnote{Data are collected every 2.5 seconds and compiled into daily reports,
available at http://weather.namsearch.com.%
} between Apr 2009 and May 2011, one day (12/7/2009) of snow was reported
and one day of fog (24/6/2009), which corresponds to that given by
Figure \ref{fig:Days-of-fog}. 5\% of the days some rain was reported,
with 71~mm total rain in 2010, similar to that given by Figure \ref{fig:Rain}.

The average wind speed was 16~km/h, with a maximum speed of 80~km/h.
Only 6\% of days had a maximum wind speed larger than 50~km/h, which
implies that the wind speed is far below 50~km/h for more than 94\%
of the time.

The difference in minimum and maximum temperatures each day is about
15 degrees Celsius, with the temperatures reaching zero in winter
and 35 degrees Celsius in summer as shown in Figure \ref{fig:Aus_Temp}.
As expected from the cold temperatures, there are some days of frost
as shown in Figure \ref{fig:Days-of-frost}.

\section{Accessibility \& infrastructure}

Although the Kuibis sites are relatively close to the B4 national
highway (Figure \ref{fig:B4}), only the areas 2b and 2c is easy to
access from the B4 (see Figures \ref{fig:contour_2b} and \ref{fig:contour_2c}).
The area 2d can be accessed by a good gravel road from Bethanie about
60~km away (see Figure \ref{fig:Namibian3}). Currently there is
a farm road (about 10~km long) to area 2a from the same Bethanie
gravel road. It would be possible to access area 2a from the B4 (via
area 2b) as well, but it is not clear if such a road already exists.

Using area 2b as reference, the area is about 60~km from the railway
town Aus (population of 300) and 160~km from the harbor town Luderitz
(population of 30k) west along the B4. The area is 60~km from Bethanie
(population of 10k) and 160~km from the provincial capital Keetmanshoop
(population of 25k) east along the B4. 

\textbf{Accommodation, gasoline, food and general supplies} are available
at both Aus and Bethanie. The nearest government \textbf{hospitals}
and \textbf{fire stations} are at Keetmanshoop and Luderitz, although
Bethanie also offers some medical services at its clinic. Both Keetmanshoop
and Luderitz have hospitals with \textbf{medical emergency services}
and ambulances that can reach the sites within 2 hours. Two medical
air rescue services, E-med Rescue and International SOS, operate from
Windhoek and can land at Aus, Keetmanshoop or Luderitz to transfer
patients to either Windhoek or South-Africa.

Aus has an \textbf{air landing strip} and Bethanie has a \textbf{small
airport}. Luderitz has the second largest \textbf{harbour} in Namibia
(after Walvis Bay) and also a \textbf{national airport} with commercial
flights (operated by Bay Air) from Windhoek three times a week. (Air
Namibia stopped scheduled flights to Luderitz from Sept 2010.) 

Keetmanshoop has an international airport with customs, although currently
there seems to be no commercial flights to Keetmanshoop. Via Keetmanshoop,
the site is about 660~km from the capital Windhoek, which has a large
\textbf{international airport}. There are also bus services connecting
Keetmanshoop to Windhoek, Cape Town, Bethanie, Aus and Luderitz at
least once a day. \textbf{Rental car} facilities are available in
Luderitz and Keetmanshoop and Keetmanshoop has a shuttle service (ideal
for groups of people).

The major \textbf{railway} line between Luderitz and Keetmanshoop
approximately follows the B4, with the nearest \textbf{railway} \textbf{station}
at Aus and the nearest siding at Guibes (20~km from area 2b). 

A 132 kV \textbf{power} line also follows the B4 past the site, with
distribution stations at Aus and Goageb. The area has clean drinking
\textbf{water} that is supplied by boreholes.

The southern backbone ring network of Namibia passes through Aus with
a STM-64 connection from Aus to Keetmanshoop%
\footnote{Telecom Namibia 2008/9 Annual report.%
} which probably also follows the B4 past the site. \textbf{High bandwidth
connections} are therefore possible from Aus.

\theendnotes

\appendix

\chapter{Contour maps}

\begin{figure}[h]
\noindent \begin{centering}
\includegraphics[scale=0.4]{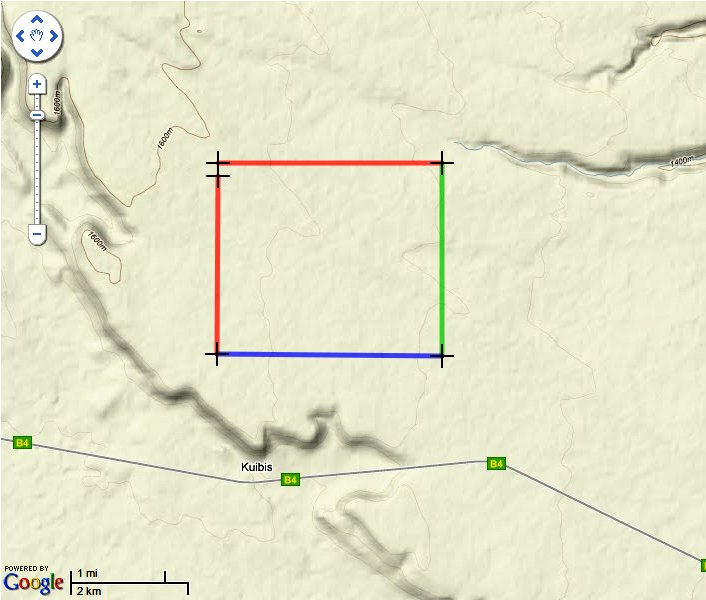}
\par\end{centering}

\noindent \begin{centering}
\includegraphics[width=1\columnwidth]{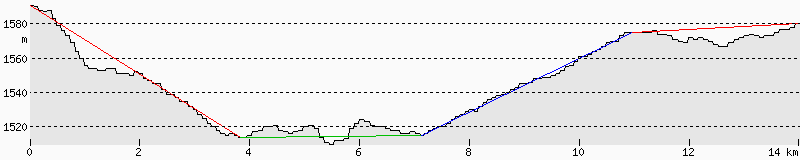}
\par\end{centering}

\caption{Site 2b: 40~m contour map with path profile around selected area\label{fig:contour_2b}}

\end{figure}
\begin{figure}
\noindent \begin{centering}
\includegraphics[width=1\columnwidth]{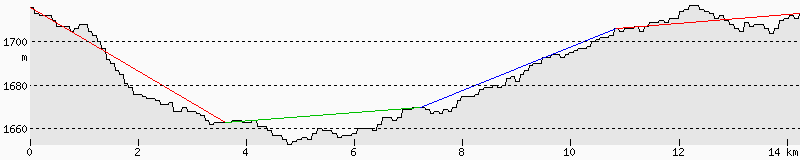}
\par\end{centering}

\caption{Site 2a: Path profile around selected area}

\end{figure}

\begin{sidewaysfigure}
\noindent \begin{centering}
\includegraphics[scale=0.4]{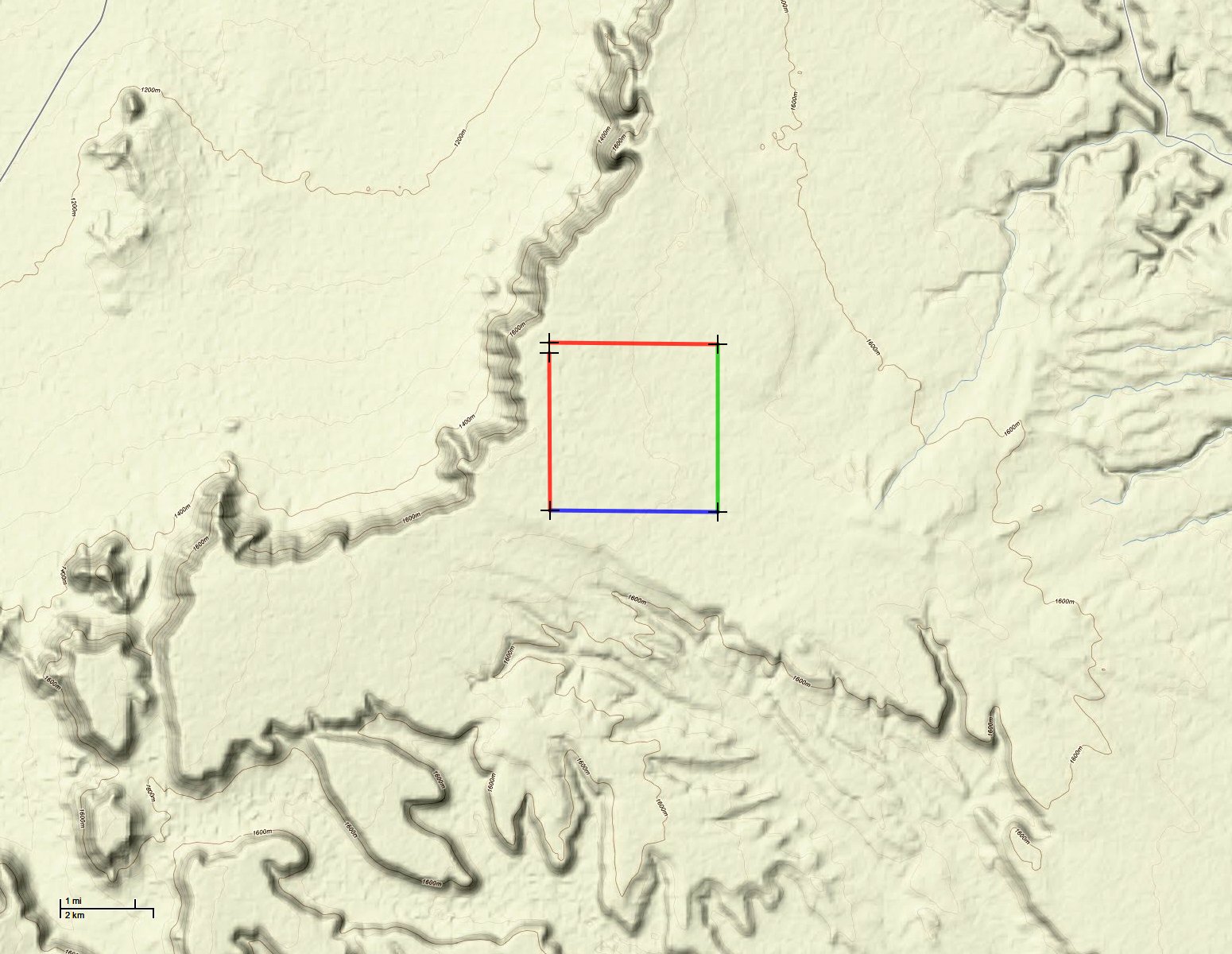}
\par\end{centering}

\caption{Site 2a: 40~m contour map\label{fig:contour_2a}}

\end{sidewaysfigure}
\begin{figure}
\noindent \begin{centering}
\includegraphics[scale=0.4]{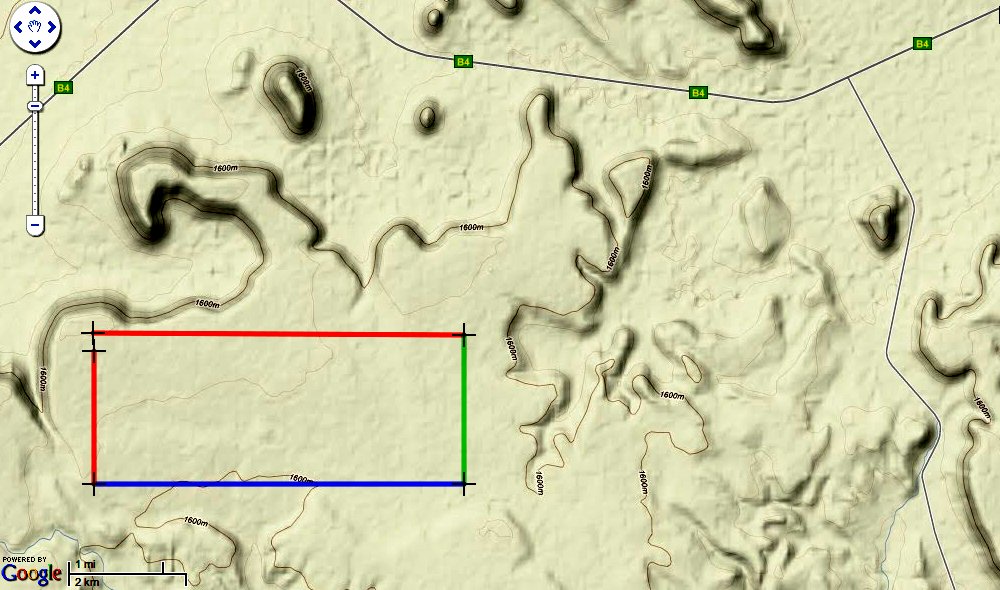}
\par\end{centering}

\noindent \begin{centering}
\includegraphics[width=1\columnwidth]{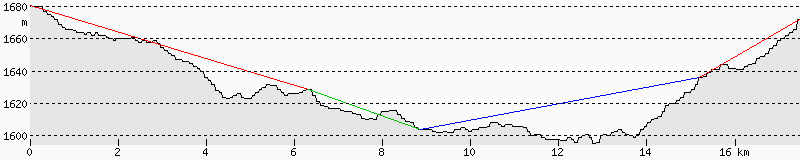}
\par\end{centering}

\caption{Site 2c: 40~m contour map with path profile around selected area\label{fig:contour_2c}}

\end{figure}
\begin{figure}
\noindent \begin{centering}
\includegraphics[scale=0.4]{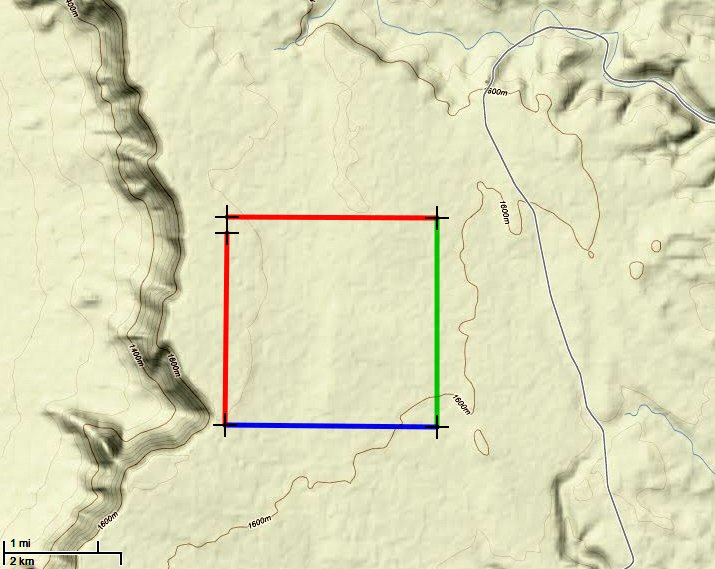}
\par\end{centering}

\noindent \begin{centering}
\includegraphics[width=1\columnwidth]{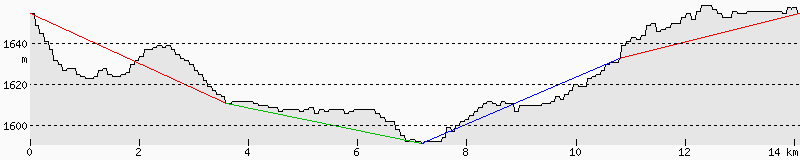}
\par\end{centering}

\caption{Site 2d: 40~m contour map with path profile around selected area\label{fig:contour_2d}}

\end{figure}

\begin{figure}[h]
\noindent \begin{centering}
\includegraphics[scale=0.4]{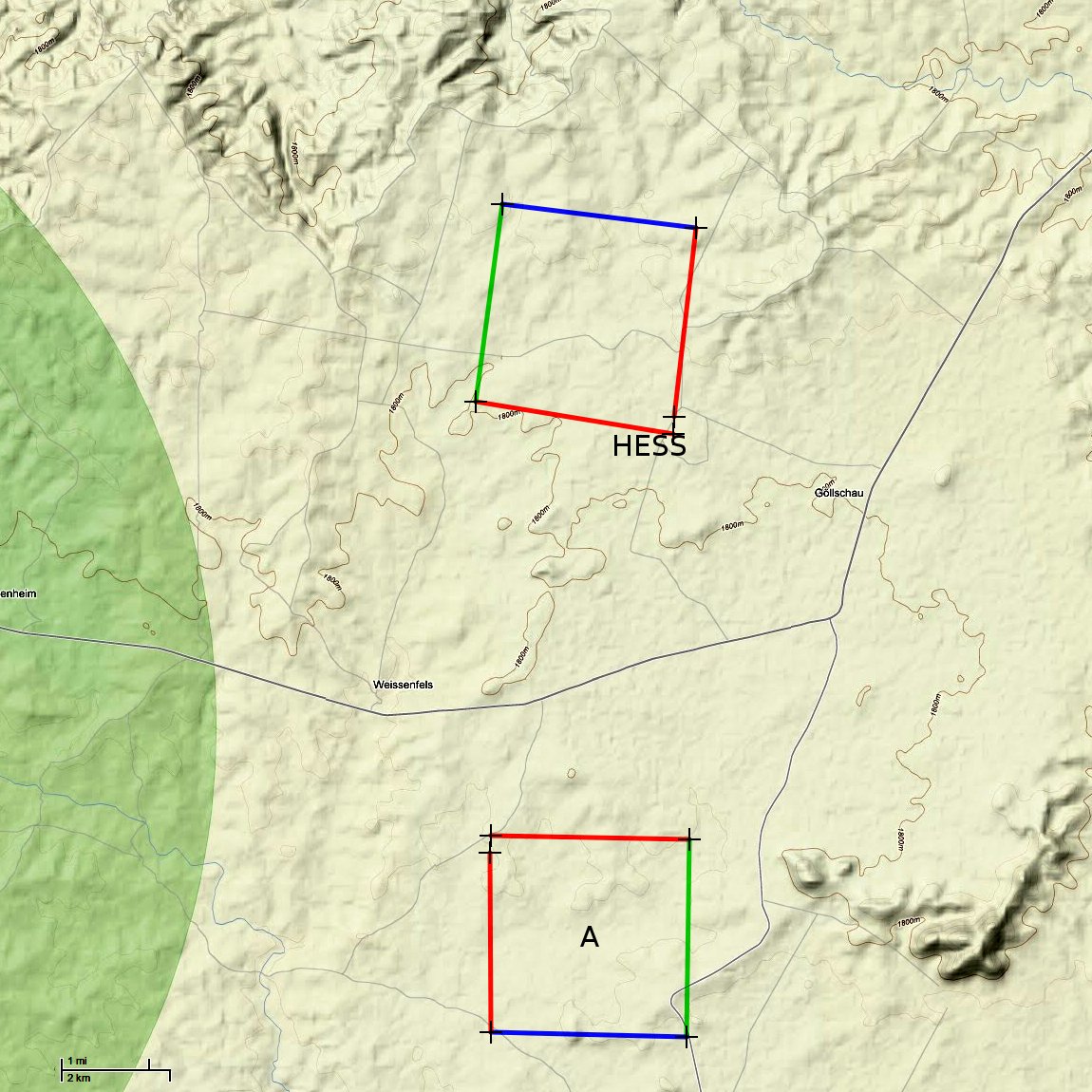}\\
H.E.S.S.\\
\includegraphics[width=1\columnwidth]{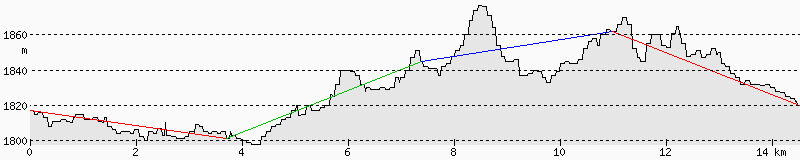}\\
A\\
\includegraphics[width=1\columnwidth]{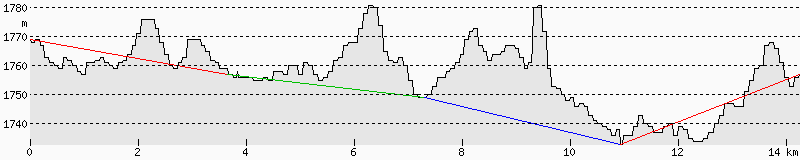}
\par\end{centering}

\caption{Site 4: 40~m contour map with path profile around selected area\label{fig:contour_4}}

\end{figure}

\begin{sidewaysfigure}
\noindent \begin{centering}
\includegraphics[scale=0.4]{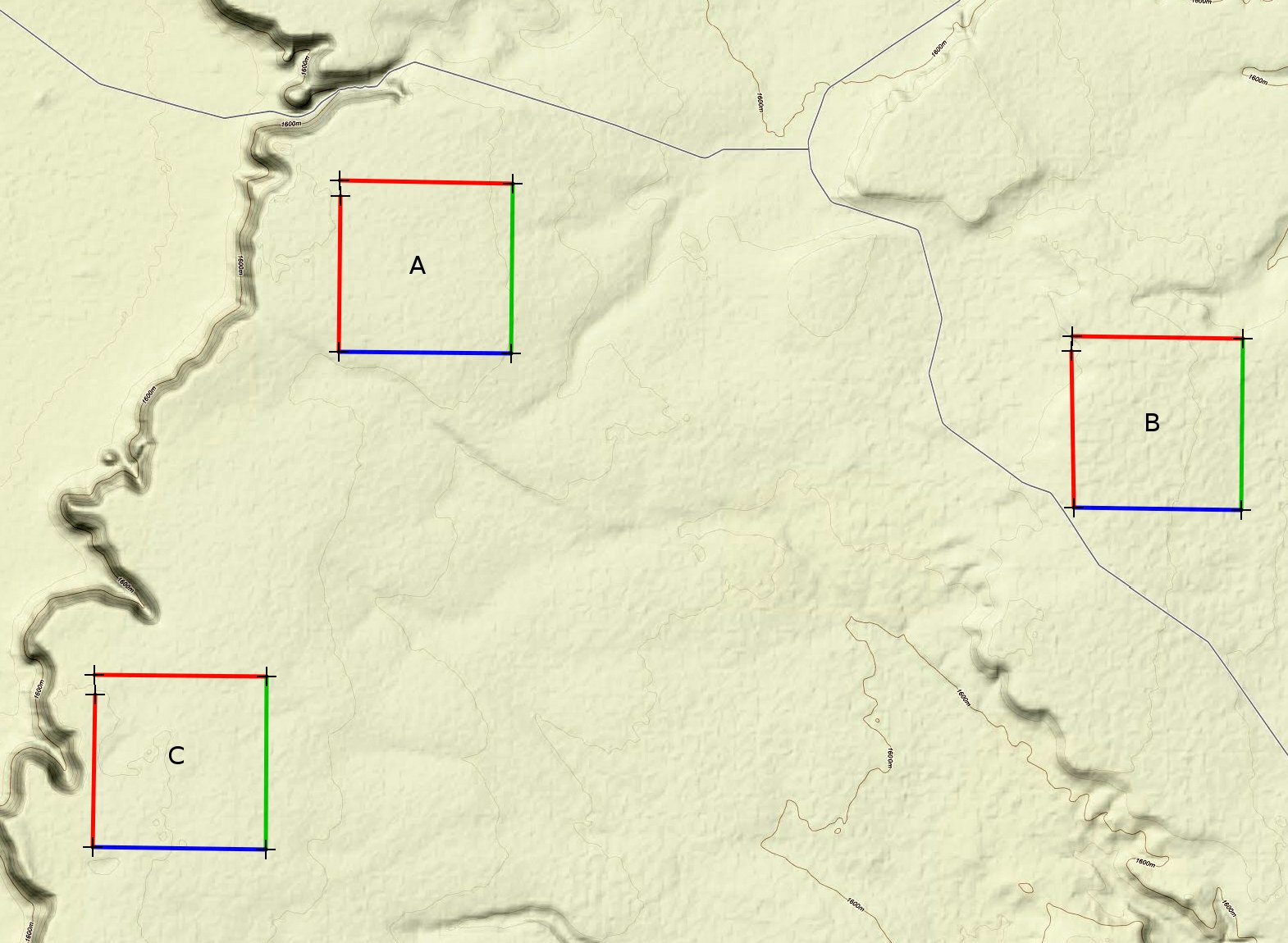}
\par\end{centering}

\caption{Site 3: 40~m contour map\label{fig:contour_3}}

\end{sidewaysfigure}

\begin{figure}[h]
\noindent \begin{centering}
A:
\par\end{centering}

\noindent \begin{centering}
\includegraphics[width=1\columnwidth]{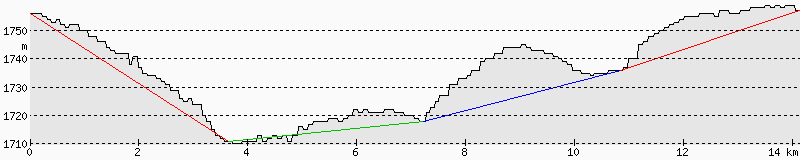}
\par\end{centering}

\noindent \begin{centering}
B
\par\end{centering}

\noindent \begin{centering}
\includegraphics[width=1\columnwidth]{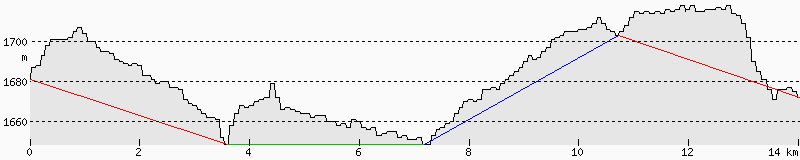}
\par\end{centering}

\noindent \begin{centering}
C
\par\end{centering}

\noindent \begin{centering}
\includegraphics[width=1\columnwidth]{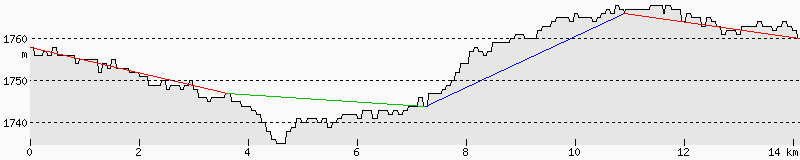}
\par\end{centering}

\caption{Site 3: Path profile around selected areas}

\end{figure}

\chapter{Political and economic overview}

\section{World Bank Country Brief: South Africa}

South Africa\textquoteright{}s 1994 transition from apartheid to constitutional
democracy remains one of the most important and impressive political
transitions of our time. It is a powerful demonstration of the proposition
that a peaceful, negotiated path from conflict and injustice to cooperation
and reconciliation is possible, despite the complex history of oppression,
institutionalized violence, and diverse social fabric that has defined
South Africa.

Since 1994, the African National Congress (ANC) has won landslide
victories in all four democratic elections. Elections are well-managed
and fair, and the press is unrestrained. Opposition parties, among
them the Democratic Alliance, Independent Democrats, and the Congress
of the People (COPE) enjoy full political freedoms.

In April 2009, the country held its fourth general elections, and
predictably the ANC won obtaining a 65.9 percent majority. H.E. Jacob
Zuma was sworn in as president of South Africa in May 2009. A new
cabinet was announced as a tripartite alliance that includes the Congress
of South African Trade Unions (COSATU) and the South African Communist
Party (SACP). The cabinet was expanded from 28 to 34 ministries to
include improved planning, performance monitoring, and service delivery
functions.

South Africa is increasingly gaining prominence on the international
stage where it is becoming a more active participant in events such
as the Annual Meetings of the International Monetary Fund and the
World Bank, the G-20, and the G-24. South Africa\textquoteright{}s
tremendous success in hosting the 2010 World Cup has also helped to
burnish its image globally. President Zuma has also made outreach
to emerging economies and alliances such as BRIC and IBSA as a priority
for his presidency. South Africa will host the 2011 meeting of the
Convention of Parties, U.N. Framework Convention on Climate Change
in November and December 2011. South Africa continues to grapple with
extreme differences in incomes and wealth. Robust economic growth
in the post-apartheid period has enabled a measurable decline in income
poverty. However, inequality has increased and as measured by the
Gini coefficient, inequality rose from 0.64 to 0.67 in the period
1995 to 2008. At over 25 percent, the unemployment rate remains very
high, and the poor continue to have limited access to economic opportunities
and basic services.

Human development challenges also loom large. These include a disappointingly
low life expectancy of 51 years. South Africa has the largest number
of people living with HIV/AIDS in the world (over 5.5 million) and
continues to battle a dual epidemic of tuberculosis and HIV/AIDS,
bearing 24 percent of the global burden of HIV-related tuberculosis.
Current health problems are rooted in the unique legacy of South African
apartheid history. The migrant labor system has contributed to many
of the major current health problems through social changes that have
led to destruction of family life, alcohol abuse, and violence\textemdash{}particularly
gender-based violence\textemdash{}while the health delivery system
is heavily skewed in favor of the elite. 

At the same time, policies working to rectify the many inequities
of apartheid have contributed to some notable achievements. Between
1991 and 2008, net secondary school enrolment went up from 45 percent
to 72 percent. Since the end of apartheid, 1.6 million free housing
units have been constructed for low-income families. Access to electricity
went up from 34 percent in 1993 to 81 percent in 2007. Similar improvements
have also been recorded for water and sanitation. The social grant
system, which primarily supports children, has expanded coverage from
2.5 million beneficiaries in 1999 to over 13 million in 2009.

\subsection{Recent Economic and Social Developments}

Prior to the financial crisis South Africa\textquoteright{}s economic
policy was largely successful despite underlying structural issues
including high unemployment, low domestic savings and investment,
and a large current account deficit. The global economic downturn
helped trigger South Africa\textquoteright{}s first recession in 17
years. However, as 2010 draws to a close, a broad-based upturn in
the economy has prompted a 3 percent annual growth rate. Driven primarily
by renewed global demand for commodities and spending related to the
World Cup 2010, the turnaround has also been supported by a revival
in the automobile industry as well as increased demand for chemical
products.

The fiscal space created by years of budgetary discipline has also
played a role in the turnaround. South Africa\textquoteright{}s low
public debt combined with its deep and liquid capital markets have
provided the access to global finance necessary for the government
to expand its own spending in areas such as infrastructure and social
services. Although this spending has resulted in the national budget
going from a surplus of 1.7 percent of GDP in 2007 and 2008 to a deficit
of 7.3 percent of GDP in 2009 and 2010, international trade and the
financial sector remain strong. After an initial dip following the
onset of the financial crisis, foreign portfolio investment has recovered
and provides solid cover for South Africa\textquoteright{}s large
current account deficit. There has also been an increase in foreign
trade with imports outpacing exports.

Throughout 2010 these underlying socioeconomic tensions came to a
head in the form of industrial action and lightning strikes. In August,
a coalition of unions representing over 1 million public servants\textemdash{}including
teachers, doctors, nurses, police, court officials, and government
bureaucrats\textemdash{}launched a strike after the government rejected
the unions\textquoteright{} demand for an 8.6 percent salary increase
and 1,000 rand monthly housing allowance. Pressures were exacerbated
by the fact that in order to avoid strikes during the World Cup, public
entities such as Eskom and Transnet had approved similar wage increases
earlier in the year. As a result, perception of widening disparities
between haves and have-nots has become acute, jumping to the forefront
of the national psyche and fueling discontent about poverty, lack
of employment opportunities, and rampant failures in service delivery
especially for poor people. Since late 2008, the ranks of \textquotedblleft{}discouraged
workers\textquotedblright{} have swelled by 739,000. The disenchantment
has been simultaneously compounded by South Africa\textquoteright{}s
difficult macroeconomic situation, and although the economy is beginning
to show signs of growth, employment generation continues to lag. Furthermore,
jobs shed in aftermath of the global financial crisis which hit South
Africa harder than other middle income countries in Eastern Europe
and Asia, are yet to be replaced exacerbating an already dire unemployment
situation.

\subsection{Government Policy priorities}

South Africa\textquoteright{}s development strategy faces a number
of significant challenges, including accelerating growth and sharing
its benefits more broadly, extending opportunities to all and improving
the coverage of delivery of public services.

To help address these challenges, the South African government has
launched a Medium Term Strategic Framework (MTSF) for 2009 to 2014
with ten priorities: 
\begin{itemize}
\item more inclusive economic growth, decent work and sustainable livelihoods 
\item economic and social infrastructure 
\item rural development, food security and land reform 
\item access to quality education 
\item improved health care 
\item the fight against crime and corruption 
\item cohesive and sustainable communities 
\item improving public service delivery 
\item sustainable resource management and use 
\item support for the creation of a better Africa.
\end{itemize}
The Government of South Africa has placed particular emphasis on accountability
for quality of outcomes, and has recently introduced performance contracts
for cabinet ministers.

\emph{(http://go.worldbank.org/GSBYF92330 as last updated on September
2010)}

\section{World Bank Country Brief: Namibia}

Namibia is a middle-income country whose considerable successes rest
on a strong multiparty parliamentary democracy that delivers sound
economic management, good governance, basic civic freedoms, and respect
for human rights. At independence in 1990, Namibia inherited a well-functioning
physical infrastructure, a market economy, rich mineral resources,
and a relatively strong public administration. However, the social
and economic imbalances of the former apartheid system have left Namibia
with a highly dualistic society. The structure of the economy has
made job creation difficult, and poverty and inequality remain unacceptably
high. These key challenges are at the top of the government\textquoteright{}s
development agenda.

Namibia has made significant progress in addressing the structural
problems: access to basic education has become more equitable and
primary health care services are widely available. Access to safe
water and sanitation has improved, and sound public policies are helping
to lay the foundation for gender parity and new programs have been
launched to protect the country\textquoteright{}s environment and
natural resources. Namibia not only maintains a social safety net
for the elderly, disabled, orphans, vulnerable children, and war veterans,
but also has a Social Security Act that provides for maternity leave,
sick leave, and medical benefits to the population.

Nonetheless, human development challenges persist. Namibia is ranked
128 out of 182 countries surveyed in the 2009 Human Development Report.
Although Namibia is on track to meet the Millennium Development Goals
on education, environment and gender, it faces daunting challenges
in combating the HIV/AIDS epidemic, making it especially challenging
to meet Millennium Development Goal Six.

\subsection{Economic overview}

A politically stable upper middle-income country with a per capita
income of approximately US\$4,310, Namibia\textquoteright{}s economy
is closely linked to South Africa\textquoteright{}s economy, and the
Namibian dollar is pegged to the South African rand. As a result,
economic trends including inflation closely follow those in South
Africa. Prior to the 2009 global financial crisis, Namibia had experienced
steady growth, moderate inflation, limited fiscal debt, a robust mining
sector, a fairly developed infrastructure, and a strong legal and
regulatory environment. From 1990 to 2008, economic growth averaged
4.5 percent per year.

However, the 2009 onset of the global economic crisis has not only
lowered demand for Namibia\textquoteright{}s commodity exports, mainly
diamonds, but also reduced the transfer payments the country receives
due to its membership in the Southern African Customs Union (SACU).
Following years of successive growth, the Namibian economy recorded
a negative growth of 0.8 percent in 2009. Following three consecutive
years of budget surpluses, the government responded to the sudden
economic downturn by running a budget deficit, and fiscal deficits
are expected to widen. For the next three years, the International
Monetary Fund (IMF) has projected deficits of 8.1 percent, 7.8 percent,
and 3.8 percent of GDP respectively. In 2009 and 2010, public debt
stood at 14.9 percent of GDP, well below the government\textquoteright{}s
target of 25 percent. Nevertheless, in 2010 the economy has shown
signs of a significant rebound due to government investment and rising
commodity exports, and economic forecasters are now predicting a growth
rate of 4.4 percent.

\subsection{Political Context}

The presidential and National Assembly elections that took place in
late November 2009 confirmed the dominance of the South West Africa
People's Organization (SWAPO) party in national politics, achieving
an overwhelming victory. The newly-formed Rally for Democracy and
Progress Party, headed by the former SWAPO Party Minister of Trade
and Industry, Mr. Hidipo Hamutenya, became the official opposition
overtaking the Congress of Democrats Party.

In the 2009 election manifesto, the SWAPO Party ran on the commitment
that it would work to \textquotedblleft{}Consolidate monetary and
fiscal policies geared towards promoting growth and employment creation,
and to ensure forward and backward economic linkages of economic sectors
and regions.\textquotedblright{} Furthermore, the party promised to
\textquotedblleft{}maintain prudent macroeconomic policies that are
responsive to domestic, regional and international economic developments.\textquotedblright{}

\subsection{Development Challenges}

Although Namibia has sustained a noteworthy track record on economic
growth and macroeconomic stabilization, certain daunting development
challenges remain. In particular, while poverty rates have declined
since independence, widespread unemployment and distribution of income
and assets remain significant issues. With a Gini coefficient of 0.74
(UNDP, 2007), Namibia is among some of the least equitable countries
in the world. Thus, a central policy challenge in Namibia is to achieve
higher rates of growth, create jobs, alleviate poverty, reduce inequality,
and raise living standards.

Similarly, despite a decline in HIV prevalence rates, which have fallen
from 22 percent in 2002 to 17.8 percent in 2008, HIV infections remain
a serious concern. Namibia also has one of the highest tuberculosis
prevalence rates in the world at 765 per 100,000 with some regions
reporting tuberculosis rates as high as 1,000 per 100,000.

In addition, Namibia is expected to face increased budgetary constraints
in coming years as SACU transfer payments are expected to continue
their downward trend. 

\emph{(http://go.worldbank.org/1B6KN88H10 as last updated on September
2010)}

\section{Economic overview}

The Institute for Security Studies (ISS) is a pan-African applied
policy research institute headquartered in Pretoria, South Africa
with the objective to add critical balance and objectivity by providing
timely, empirical research and contextual analysis of relevant human
security issues to policy makers, area specialists, advocacy groups,
and the media. ISS provide the following overview of the South Africa
and Namibian economies:
\begin{quote}
{}``\textbf{South Africa} is a middle-income, developing country
with an abundant supply of resources, well-developed financial, legal,
communications, energy, and transport sectors, a stock exchange that
ranks among the largest ten in the world, and a modern infrastructure.
However, high levels of unemployment (around 40\% in 2000) and economic
problems that are a legacy of the apartheid, notably poverty and the
lack of economic empowerment among the disadvantaged groups, are serious
structural problems, South Africa having one of the most unequal distributions
of income in the world. Despite the government's commitment to promote
job creation in accordance with its Growth, Employment and Redistribution
(GEAR) strategy, and the number of jobs declined by some 600,000 between
1995-98. Economic growth continues to lag behind population growth.
Other pressing problems are crime and the high incidence of HIV/AIDS.
With 20 per cent of the adult population estimated to be HIV positive,
the incidence of AIDS presents a major threat to the country's long-term
skills base.

Economic activity is centered on the minerals and energy sector with
manufacturing based mainly on mining activity. Exports are driven
mainly by mining, energy and agriculture-related activities, with
gold remaining the largest foreign-exchange earner (some 30\% of exports),
and platinum the country\textquoteright{}s chief export. South Africa
is the world\textquoteright{}s biggest platinum supplier, and it expected
to provide 74\% of global market requirements in 2000. The service
sector is the most important contributor to GDP at 57\% (1998). It
ranges from the financial sector to a developing tourism and an important
retail sector. Agriculture contributes little to GDP, but is linked
to significant agro-industrial activities. Informal services are for
many the main source of employment. Inflation fell from 8.5 per cent
in 1997 to 5.5 per cent in 1999, rising to 7.2\% in 2000. Inflation
is expected to average 6\% in 2001.'' (http://www.iss.co.za/af/profiles/SouthAfrica/southafrica1.html)

{}``\textbf{Namibia}'s economy is closely tied to that of South Africa
through a number of institutional relationship's - in particular the
Southern African Custom's Union and the Common Monetary Area - as
well as through extensive trade and financial flows. Namibia has abundant
natural resources, good infrastructure and access to markets, but
contrary to potential, the economy is not well diversified. Economic
activity is concentrated in primary sector activities - the extraction
and processing of minerals for export which accounts for 20\% of GDP,
large scale commercial livestock farming, and fishing. The economy
is highly vulnerable to world market price fluctuations of diamonds
and uranium, prices and demand remaining crucial to the country\textquoteright{}s
economic prospects. Government services accounted for some 25\% of
GDP in 2000.

While Namibia appears to be a prosperous, middle-income country, the
estimated GDP per capita of US\$ 1,800 camouflages huge income disparities,
presenting Namibia one its main economic problems. The majority of
the population is poor, with limited access to social services in
certain areas. Furthermore, the historic pattern of land ownership
means the majority of Namibians are landless or small-scale subsistence
farmers.'' (http://www.iss.co.za/af/profiles/Namibia/Economy.html)
\end{quote}
\textbf{Euler Hermes Country Risk Grade} for South-Africa is BB and
Namibia is B (on the scale AA, A, BB, B, C, D with AA the best). The
report gives the following strengths, weaknesses and risks for South
Africa and Namibia:

\subsubsection*{South Africa's Euler Hermes Country Risk Profile}

\begin{minipage}[t]{0.45\columnwidth}%
\begin{itemize}
\item Strengths 

\begin{itemize}
\item Geographic and economic size engender regional dominance 
\item No external threats to sovereignty 
\item ANC government has a strong mandate; political stability to continue 
\item Political, judicial and security systems entrenched and aligned with
Western `norms' 
\item Economic management is good, with monetary and fiscal policy agencies
highly regarded 
\item Exemplary exodus from foreign debt problems of the mid-1980s 
\item Business environment generally favourable, with practices familiar
to European practitioners 
\item Relations with IFIs are good 
\end{itemize}
\end{itemize}
\end{minipage}%
\begin{minipage}[t]{0.45\columnwidth}%
\begin{itemize}
\item Weaknesses 

\begin{itemize}
\item Open economy leads to occasional currency and external account pressures 
\item Long-term problems require policy and action, including unemployment,
rural poverty and HIV/AIDS (an estimated 5.7mn are infected out of
a total population of around 50mn) 
\item Lingering concerns in relation to LT policy stance now that Jacob
Zuma is head of state 
\item Current account and fiscal deficits require careful management 
\item Danger of policy ossification in regard to privatisation and labour
relations 
\item Inward investment weighted towards portfolio flows rather than FDI 
\item Power generation has not increased sufficiently to meet growing demand
and there have been periods of rationing of supply for homes and industry 
\item Relatively weak liquidity levels 
\item Labour market inflexibility
\end{itemize}
\end{itemize}
\end{minipage}
\begin{itemize}
\item Key Risks 

\begin{itemize}
\item The rand is an openly traded currency and is susceptible to volatility.
An external event causing weakness in emerging market currencies as
a class could lead to a run on the rand, with associated policy concerns
and deterioration in economic data 
\end{itemize}
\end{itemize}

\subsubsection*{Namibia's Euler Hermes Country Risk Profile}

\begin{minipage}[t]{0.45\columnwidth}%
\begin{itemize}
\item Strengths\textbf{ }

\begin{itemize}
\item Stable democracy since independence in 1990 
\item Political environment has matured as indicated by the successful transition
from long-serving Sam Nujoma to President Hifikepunye Pohamba 
\item Close association with South Africa through customs and monetary unions 
\item Natural resource base includes substantial diamonds (including offshore) 
\item Current account surpluses 
\item Manageable foreign debt ratios 
\end{itemize}
\end{itemize}
\end{minipage}%
\begin{minipage}[t]{0.45\columnwidth}%
\begin{itemize}
\item Weaknesses 

\begin{itemize}
\item Land reform programme is a policy priority and there is concern that
it may yet develop into a Zimbabwe-style land -grab 
\item Diversification of economy is relatively limited 
\item Small population base and limited arable land area 
\item Fiscal deficits 
\item Poverty and unemployment remain high 
\item Foreign reserves holdings, though improving, provide limited import
cover 
\end{itemize}
\end{itemize}
\end{minipage}
\begin{itemize}
\item Key Risks 

\begin{itemize}
\item The economy is strongly linked to that of South Africa and a rapid
deterioration in the latter would have negative repercussions in Namibia 
\item The high prevalence of HIV/AIDS in a country with a small population
limits the available manpower and the economy, in general, and the
non-mining sector, in particular, are therefore unlikely to reach
potential 
\end{itemize}
\end{itemize}


\begin{thebibliography}{9} \providecommand{\natexlab}[1]{#1}

\bibitem[{Bulik \emph{et~al.}(2009)Bulik, Cieslar, Boisson, Median \&   team}]{Bulik2009} \textsc{Bulik, T., Cieslar, M., Boisson, C., Median, C. \& team, S.W.} 2009. \newblock Automated CTA site search -- initial report.

\bibitem[{Centre(2008)}]{Centre2008} \textsc{Centre, N.I.} 2008. \newblock Cost of doing business in Namibia. \newblock Ministry of Trade \& Industry.

\bibitem[{Cinzano \emph{et~al.}(2001)Cinzano, Falchi \& Elvidge}]{Cinzano2001} \textsc{Cinzano, P., Falchi, F. \& Elvidge, C.} 2001. \newblock The first world atlas of the artificial night sky brightness. \newblock \emph{Mon. Not. R. Astron. Soc.}, 328:689--707.

\bibitem[{CTA(2010)}]{CTA2010} \textsc{CTA} 2010. \newblock Call for site proposals.

\bibitem[{Forsythe \emph{et~al.}(1995)Forsythe, Rykiel, Stahl, Wu \&   Schoolfield}]{Forsythe1995} \textsc{Forsythe, W., Rykiel, E., Stahl, R., Wu, H. \& Schoolfield, R.} 1995. \newblock A model comparison for daylength as a function of latitude and day of   year. \newblock \emph{Ecological Modelling}, 80:87--95.

\bibitem[{Harding(1974)}]{Harding1974} \textsc{Harding, G.} 1974. \newblock Observing conditions at Sutherland 1972 August to 1973 December. \newblock \emph{Circulars. South AfricaN Astron. Obs.}, 1:31--37.

\bibitem[{NRF(2005)}]{NRF2005} \textsc{NRF} 2005. \newblock Proposal to site the Square Kilometre Array. \newblock National Research Foundation, South Africa.

\bibitem[{Preuss \emph{et~al.}(2002)Preuss, Herman, Hofmann \&   Kohnle}]{Preuss2002} \textsc{Preuss, S., Herman, G., Hofmann, W. \& Kohnle, A.} 2002. \newblock Study of the photon flux from the night sky at La Palma and Namibia,   in the wavelength region relevant for imaging atmospheric Cherenkov   telescopes. \newblock \emph{Nuclear Instruments and Methods in Physics Research A},   481:229--240.

\bibitem[{Venter(2011)}]{Venter2011} \textsc{Venter, C.} 2011. \newblock The status of H.E.S.S.\ and CTA. \newblock \emph{African Skies / Cieux Africains}, 16.
\end{thebibliography}
\end{document}